\documentclass[journal]{IEEEtran}

\usepackage{xcolor}

\usepackage{graphicx} 
\usepackage{hyperref}
\usepackage{epstopdf}
\usepackage{gb4e}
\usepackage{cite}
\usepackage{amsmath}	

\graphicspath{{images/}}

\begin{document}
		%
		\title{Unveiling Malware Patterns: A Self-analysis Perspective}
		%
		%
		%
		
		\author{Fangtian~Zhong,~\IEEEmembership{Member,~IEEE,}
			 Qin~Hu,~\IEEEmembership{Member,~IEEE,}
        Yili~Jiang,~\IEEEmembership{Member,~IEEE,}
			Jiaqi~Huang, ~\IEEEmembership{Member,~IEEE,}
      
  and~Xiuzhen~Cheng,~\IEEEmembership{Fellow,~IEEE}
			\thanks{Fangtian~Zhong is with the Gianforte School of Computing,
				 Montana State University, Bozeman,
				MT 59718, USA. E-mail: fangtian.zhong@montana.edu}
			\thanks{Qin~Hu and Yili~Jiang are with the Department of Computer Science, Georgia State University, Atlanta, GA 30303, USA. E-mail: \{qhu, yjiang27\}@gsu.edu}
                \thanks{Jiaqi~Huang is with the Department of Computer Science and Cybersecurity, University of Central Missouri, Warrensburg, MO 64093, USA. E-mail: jhuang@ucmo.edu}
            \thanks{Xiuzhen~Cheng (Corresponding author) is with the College of Computer Science and Technology, Shandong University, Qingdao, Shandong 266237, China. E-mail: xzcheng@sdu.edu.cn}
        }
		
		%
		%

	\markboth{Journal Latex}%
	{Shell \MakeLowercase{\textit{et al.}}: Bare Demo of IEEEtran.cls for IEEE Journals}
	%



	\maketitle
	
	\begin{abstract}
The widespread usage of Microsoft Windows has unfortunately led to a surge in malware, posing a serious threat to the security and privacy of millions of users. In response, the research community has mobilized, with numerous efforts dedicated to strengthening defenses against these threats. The primary goal of these techniques is to detect malicious software early, preventing attacks before any damage occurs. However, many of these methods either claim that packing has minimal impact on malware detection or fail to address the reliability of their approaches when applied to packed samples. Consequently, they are not capable of assisting victims in handling packed programs or recovering from the damages caused by untimely malware detection. In light of these challenges, we propose \textit{VisUnpack}, a static analysis-based data visualization framework for bolstering attack prevention while aiding recovery post-attack by unveiling malware patterns and offering more detailed information including both malware class and family. Our method includes unpacking packed malware programs, calculating local similarity descriptors based on basic blocks, enhancing  correlations between descriptors, and refining them by minimizing noises to obtain self-analysis descriptors. Moreover, we employ  machine learning to learn the correlations of self-analysis descriptors through architectural learning for final classification. Our comprehensive evaluation of \textit{VisUnpack} based on a freshly gathered dataset with over 27,106 samples confirms its capability in accurately classifying malware programs with a precision of 99.7\%. Additionally, \textit{VisUnpack} reveals that most antivirus products in VirusTotal can not handle packed samples properly or provide precise malware classification information. We also achieve  over 97\% space savings compared to existing data visualization based methods.
		 
	\end{abstract}
	
	\begin{IEEEkeywords}
		malware classification, malware family, variants, self-analysis.
	\end{IEEEkeywords}

	%
	\IEEEpeerreviewmaketitle

	\section{Introduction}
	\IEEEPARstart{A}{s} the dominant desktop operating system globally, Microsoft's Windows has a global market share of approximately~72\%. Meanwhile, malware targeting Windows operating systems has exhibited a consistent upward trend in both volume and sophistication, accompanied by a notable diversification in functionalities over the years. These threats range from traditional viruses and worms to more advanced trojans, adware, and potentially unwanted applications (PUAs). In addition, there has been an increase in targeted attacks that utilize advanced techniques such as polymorphism, encryption, compression, and evasion tactics. According to AV-TEST \cite{anti-test}, over 97\% of malware and PUAs are distributed via Windows OSs, highlighting the platform's vulnerability to malicious activities. To address these challenges, researchers have employed various techniques, categorized into static analysis \cite{mihai} and dynamic analysis \cite{Simone}, to detect patterns and behaviors similar to malicious programs. 

Static analysis means extracting features directly from malware binaries. For instance, it extracts header information like sections, imports, exports, resource data, and subsystem details. It also identifies Windows API calls and control flows \cite{Korczynski}, capturing their frequency, sequence, and parameters. Additionally, it filters embedded strings within the binaries, such as URLs, IP addresses, registry keys, file paths, and encryption keys \cite{gu2007bothunter}. These strings are typically obtained using disassemblers like IDA Pro \cite{peng2014x}, radare2 \cite{eom2024r2i} or angr \cite{shoshitaishvili2016state}, which allow for detailed binary dissection. Unfortunately, malware authors are also  aware of the efforts to reverse engineer their binaries and employ  binary obfuscation techniques, such as packers, to deter analysis and reverse engineering. Consequently, defeating these obfuscation techniques is a critical prerequisite for performing meaningful static analysis. However, obfuscation techniques can largely be overcome by dynamic analysis. 

Dynamic analysis runs malware and observes its behaviors, ranging from low-level binary code to system-wide actions such as registry and file system modifications. Techniques such as hard-coded tests, random fuzzing, and concolic testing are employed to provoke malicious behaviors \cite{SAVIOR,ConcolicTesting}. However, relying solely on predefined inputs has a low success rate in triggering such behaviors and can be easily evaded by knowledgeable adversaries. Random fuzzing may produce redundant inputs that yield similar outcomes, while concolic testing often faces large computation and memory cost issues due to path explosions. Dynamic analysis typically employs emulators or virtual machines to create isolated environments, or bare-metal setups for a true runtime environment. Although virtual machines or emulators provide a more trusted guest-host separation, they also present challenges in bridging the semantic gap, making them vulnerable to be detected by malware. Moreover, bare-metal setups are unsuitable for analyzing malware with elevated privileges beyond user mode.  Consequently, dynamic analysis is only effective for analyzing smaller sets of malware programs. Besides, dynamic analysis typically focuses on malware detection, which may not adequately assess the extent of damage to computer systems.


Researchers have increasingly turned to artificial intelligence-driven algorithms to automatically learn features from analysis techniques or to abstract useful features from neural network embeddings for malware classification, as demonstrated by studies like \cite{Jiancong,Cuiying}. A significant focus in this field has been the use of data visualization techniques for malware classification. However, these methods classify malware programs based only on their broader classes or families and often neglect program semantics by processing the programs as a whole, as highlighted in works like \cite{Xiang,Xusheng}. Additionally, these approaches commonly assume that packing has negligible impact on classification accuracy or rely solely on unpacked datasets. As a result, their solutions lead to a lack of space savings and an underestimate of the challenges introduced by code packing. To address these challenges, we propose \textit{VisUnpack}, a classification framework that integrates static analysis, data visualization, and machine learning. \textit{VisUnpack} effectively handles the complexities introduced by code packing, distinguishes between different malware classes and families simultaneously, and optimizes the space efficiency in malware detection—an issue that other methods often struggle to resolve. Our method emphasizes understanding malware similarity through static analysis and data visualization to identify recurring patterns for more accurate classification. We evaluate the performance of \textit{VisUnpack} using a newly collected dataset, meticulously verified by the authors. This verification involved utilizing results from over 70 third-party anti-virus products on VirusTotal \cite{virustotal}, complemented by dynamic analysis, reverse engineering techniques, and the calculation of Cohen's Kappa coefficient \cite{vieira2010cohen}. It is important to note that \textit{VisUnpack} is designed to complement, rather than replace existing dynamic behavior-based systems. It enhances the efficiency of malware analysis and provides broader coverage of malware programs, particularly considering that 85\% of malware received by companies belongs to known classes or families \cite{dataset}. Our contributions can be summarized as follows:

\begin{itemize}
		\item We propose \textit{VisUnpack}, a context-sensitive malware visualization classification framework, designed to effectively distinguish between various malware classes and their variants while maintaining high accuracy. \textit{VisUnpack} achieves precision of 99.7\% while also reducing the space consumption by over 97\%. Moreover, VisUnpack ensures the preservation of similarities among malware programs within the same category and provides visualizations of self-analysis maps to aid in analysis. 

		\item  We compile a new malware dataset comprising 27,106 programs and conduct dynamic analysis, reverse engineering, and cross-validation with VirusTotal results. We find that most antivirus products on VirusTotal cannot simultaneously distinguish between malware classes and families. Consequently, methods that depend solely on VirusTotal may not scale effectively to mitigate malware.

		\item Our method thoroughly analyzes the impact of packing on classification outcomes and quantifies the degree of influence. Our results indicate that certain packers and packing strategies significantly affect classification accuracy. We also find that most antivirus products on VirusTotal fail to handle packed programs properly. Thus, our research could encourage greater interest in unpacking packed programs and developing more precise classification methods in anti-virus products that account for both malware classes and families, potentially aiding in the automatic mitigation of damage caused by malware attacks.

\end{itemize}
	
 The remainder of this paper is organized as follows: Section \ref{sec:relwork} provides an overview of related work. Section \ref{sec:methodology} presents the system design and implementation in detail. Section \ref{sec:setup} outlines the experimental settings. Section \ref{sec:evaluation} discusses the evaluation results and highlights the limitations of recent research. Finally, Section \ref{sec:conclusion} concludes the paper.
	
	\section{Related Work}
Techniques for malware classification based on data visualization can be broadly divided into two groups: image processing techniques and feature extraction techniques. In the following two subsections, we will discuss the related works for these approaches.

    \label{sec:relwork}
	\subsection{Advanced Image Processing for Malware Classification}
Cui \emph{et al.} \cite{Zhihua} proposed data augmentation techniques to enhance the quality of malware images by empirically adjusting parameters such as horizontal and vertical translation, rotation range,  magnification, projection transformation, random zooming, and nearest-neighbor padding. These adjustments improve the diversity and robustness of the dataset for malware classification. VisMal \cite{Malbrain} employed a modified histogram equalization algorithm to capture local patterns, such as pixel frequency distributions within specific regions of malware images. This method then uses the distributions to enhance the discernibility of malware programs by improving local contrasts of pixels and strengthening correlations between neighboring bytecode regions, facilitating more accurate classification. Nataraj \textit{et~al.} \cite{Nataraj} utilized the GIST technique \cite{Torralba, Aude}, which transforms images based on their global structural characteristics. This approach decomposes an image into its frequency and orientation attributes and applies a collection of Gabor filters to generate multiple filtered images. Texture-based features are then extracted from these filtered images, offering a compact and informative representation of the image’s spatial layout, thereby enhancing classification performance. Makandar \textit{et~al.} \cite{Makandar} introduced a wavelet-based approach for malware classification, consisting of three stages. In the pre-processing stage, wavelet is applied to denoise and enhance malware images. In the feature extraction stage, the Discrete Wavelet Transform (DWT) is employed to decompose images into four levels, extracting features such as db4 and coif5 wavelet coefficients. These features capture both frequency and time-domain characteristics of malware. Finally, the classification stage uses these extracted features to discriminate between malware classes, leveraging wavelet analysis for robust and accurate malware detection.

		\begin{figure*}[h]
		\centering
		\includegraphics[scale=0.39]{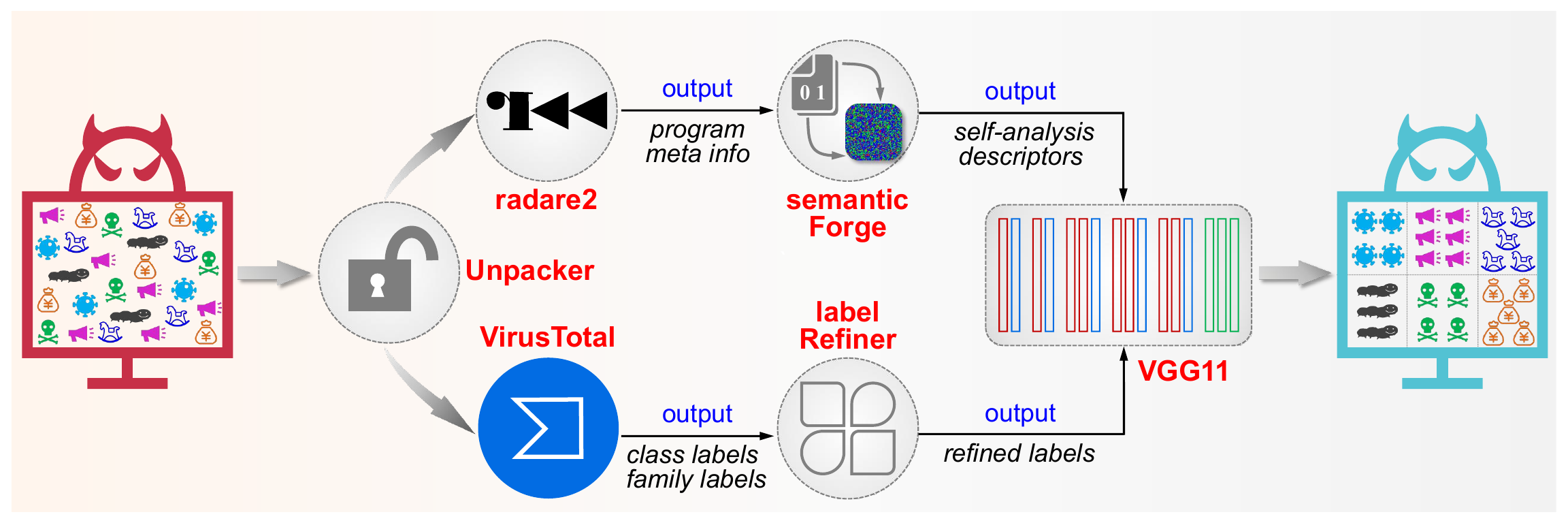}
		\caption{An Overview of VisUnpack Framework}
		\label{fig:VisUnpack}
	\end{figure*}

    \subsection{Advanced Feature Extraction Algorithms}
MDMC \cite{Baoguo} converts malware binaries into byte sequences and computes byte transfer probability matrices by analyzing the likelihood of each byte pair occurring across all possible pairs. These matrices, which capture neighboring byte correlations, are transformed into Markov images that are analyzed using a CNN to model sequential dependencies. Liu \emph{et al.} \cite{liu2019new} extract multi-scale features from malware images by dividing them into small patches and further subdividing each patch into bins. Gradient histograms are computed for local bins using a sliding window, and feature descriptors are generated via cascaded connection functions. These descriptors are organized into 16×16 sub-blocks. From the training images, 0.5\% of the sub-blocks are randomly sampled and clustered into k centers using k-means clustering. Each sub-block in a malware image is encoded based on the index of its nearest cluster center. Finally, a histogram operator processes these encoded  sub-blocks to construct the final feature vector, representing the malware image. Chen \emph{et al.} \cite{chen2020malware} disassembled malware executables into assembly code, segmented the code into basic blocks, and extracted opcodes from each block. The Simhash algorithm is then applied to these opcodes to generate hash bits for each basic block. These hash bits were subsequently converted into pixel values, where a hash bit of 0 was mapped to a pixel value of 0, and a hash bit of 1 is mapped to 255. The resulting pixels for basic blocks are concatenated to construct grayscale images, which are used for malware classification. Adkins \emph{et al.} \cite{adkins2013heuristic} adopted a similar approach but focused on abstracting features by replacing registers with “REG,” locations with “LOC,” constants with “CONST,” memory references with “MEM,” and variable references with “VAR.” Instructions are grouped into n-grams and hashed using the MD5 algorithm. The resulting MD5 hash values are truncated to the lowest 22 bits, which are used as feature hash indices. An array is then populated as the feature representation, with the value at each hash index set to 1. Other studies, such as \cite{Saxe, Ouahab, Xiang, Xusheng, Yakura}, applied machine learning algorithms directly to learn feature embeddings for malware classification.

	\subsection{Summary}
The aforementioned static analysis-based approaches, whether image processing-based or feature extraction-based, aim to achieve strong performance in general malware classification tasks, such as identifying malware classes or families. These categorizations are typically based on the behavior or techniques employed by malware. However, such behaviors can vary significantly across different variants, and the techniques may manifest differently depending on how they are implemented to infect victims. A variant refers to a new version of malware derived from existing malware with modifications. Consequently, it is crucial to focus on identifying similar code sequences, their local semantics, and their relative geometric positions. Additionally, these methods often overlook the impact of packing on malware classification or fail to address the extent to which packing affects their results, leaving this critical factor unexplored. To address these challenges, we introduce \textit{VisUnpack}, a novel system that leverages various unpackers to preprocess packed malware programs before classification. By focusing on the static features of basic blocks, \textit{VisUnpack} performs local similarity calculations and enhances the correlations between these local similarity descriptors to generate robust self-analysis descriptors for classification. Using the VGG11 model, \textit{VisUnpack} delivers a highly effective solution for accurately classifying malware programs, even across different variants of the same class or family. \textit{VisUnpack} offers several key advantages. First, it effectively circumvents obstacles in reverse analysis by unpacking malware programs. Second, it emphasizes the intrinsic characteristics of malware and uncovers underlying similarities in basic blocks, which are more resilient to changes than methods using global information. Third, by processing malware as self-analysis maps, \textit{VisUnpack} not only aids in precise classification but also significantly supports manual analysis, offering deeper insights into malware behavior and structure.\\

	\section{Design of VisUnpack}
    \label{sec:methodology}
	\subsection{Overview of VisUnpack}
	The fivefold objectives of \textit{VisUnpack} are as follows: i) resist the influence caused by packing; ii) classify malware programs into their respective classes and families; iii) preserve similarities among malware within  the same class and family while reducing the overhead on model training; iv) resist the influence of non-rigid deformation within basic blocks and between self-analysis descriptors for recognizing malware variants; v) provide a data visualization method to assist in malware analysis. 
 
 The \textit{VisUnpack} framework is presented in Fig. \ref{fig:VisUnpack}. It integrates six key components (in red font) to effectively classify malware programs. The unpacker utilizes PEid \cite{peid} to determine whether a malware sample is packed and to identify the specific packer used. It then applies the matching unpacking tool to unpack the malware program. Third-party anti-virus products, such as VirusTotal, contribute malware class and family labels, which are then refined and consolidated by the labelRefiner using methods such as regular expressions and longest common subsequences \cite{zhong2024enhancing}. Furthermore, dynamic analysis and reverse engineering are performed on each malware program, leveraging Malpedia \cite{Malpedia} and other open resources to confirm the assigned labels. The binary analysis framework, such as radare2, disassembles malware executables to instructions, identifies functions and basic blocks, and extracts essential program metadata. Following this, SemanticForge utilizes context-sensitive  techniques to extract critical local semantics within basic blocks while preserving their similarity and geo-correlation. This process generates local similarity descriptors, where the correlations between similarity descriptors are further enhanced to obtain self-analysis descriptors. Finally, the classifier uses a VGG11 model to capture local and global features in self-analysis descriptors, enabling accurate malware classification based on a hierarchy of concepts. In the following subsection, we will introduce the implementation for each component, except VGG11, as we adopt the implementation from \cite{simonyan2014very}. 
 
Once the VGG11 model is trained, it classifies malware programs through a systematic multi-step process. Initially, the malware program is analyzed to determine whether it is packed. If packing is identified, the program is unpacked using a matching unpacker. Next, the unpacked program undergoes processing with \textit{SemanticForge}, which generates self-analysis descriptors that encapsulate the program’s structural and semantic characteristics. These descriptors are then resized to a standardized input size compatible with the VGG11 model. Finally, the resized self-analysis descriptors are fed into the VGG11 model, which performs the classification.


\subsection{Key Components of VisUnpack}
\subsubsection{Unpacker}
We use PEid \cite{peid} to identify the specific packers used to obfuscate malware programs. PEid employs a database of over 5,500 signatures from sources such as app-peid, pev, ASL, MalScan, and others, enabling the detection of packers for PE files. To facilitate the unpacking process, we developed a custom script that automatically employs corresponding unpackers, such as UPX \cite{upx}, Petite \cite{petite}, PECompact\cite{PECompact}, and MPRESS \cite{mpress}, to unpack malware programs. For less common packers, we rely on two most successful generic unpackers,  Unipacker \cite{unipacker} and UnpacMe \cite{unpacme}, to ensure comprehensive unpacking of all malware programs.

\subsubsection{Third-party anti-virus products}
We rely on VirusTotal as our third-party anti-virus products. VirusTotal is a website developed by Hispasec Sistemas, a subsidiary of Google Inc. It embeds over 70 online anti-virus products to detect malware. Users can submit program files as large as 650 MB through its API by specifying the file path and a hash value calculated using SHA256, SHA1, or MD5. Upon submission, VirusTotal conducts checks using anti-virus products to identify known malware signatures, suspicious behaviors, and other indicators of compromise. The results from all anti-virus products are compiled into a comprehensive report, providing information on the file's detection status across various engines and additional details like file metadata. Due to the API limitation that allows only 500 daily submissions, it took us nearly two months to collect all labels. Currently, VirusTotal handles approximately one million submissions daily. Results are shared with antivirus vendors, who integrate new malware signatures missed by their tools into their products. Notable vendors like McAfee, F-Secure, and Microsoft contribute to VirusTotal. In our study, we utilize VirusTotal to obtain class and family labels for our label refiner process and help the construction of our dataset.

\subsubsection{labelRefiner}
 In order to avoid a single point of failure in malware recognition by antivirus products, we integrate the results from more than 70 antivirus products to assign initial labels to our dataset. Although VirusTotal aggregates results from multiple anti-virus products into a unified report for a submitted malware sample, each anti-virus product provides labels in varying formats. For example, for the same malware sample, such as $a6e197b241dc5427b6f6c170762a6732
$, Microsoft identifies it as ``PWS:Win32/VB,'' Kaspersky labels it as ``Trojan.Win32.Swisyn.bner,'' Symantec designates it as ``W32.Gosys,'' and ClamAV  detects it as ``Win.Virus.Sality:1-6335700-1.'' Due to this inconsistency, directly using these labels is impractical and may yield insufficient information. 

To address this challenge, we adopt a method inspired by \cite{zhong2024enhancing}. This approach begins by segmenting label strings using regular expressions to split them by non-alphanumeric characters, breaking each label into a set of substrings. Substrings that are not alphabetic, unrelated to malware information, or shorter than three characters are filtered out, excluding less informative elements like numeric values and very short strings. Each malware sample is then associated with a set of substrings representing potential class or family labels. While these labels may vary slightly, they convey the same underlying concept, such as ``ransom," ``ransomgen," and ``ransomkd." To refine these labels, the longest common subsequence algorithm is used to measure similarity rates and identify a suitable threshold ($thr$) for merging similar names. For labels with similarity below the threshold, a group is defined to resolve discrepancies. For instance, labels such as "pua" and "pup," both referring to potentially unwanted applications, are consolidated. To ensure consistency, labels are arranged by length, and longer ones are replaced with shorter equivalents. Next, the top $t$ most frequent substrings are identified as potential class or family labels. The substring with the highest frequency is paired with each of the remaining $t-1$ substrings to generate candidate malware class-family labels for each sample. This process prioritizes specific family labels over broad categories, such as replacing ``trojan.adware" with a more precise family label when available. This ensures finer granularity in describing the unique techniques used by malware, avoiding overly generic labels like ``trojan," ``adware," or ``pua." After generating candidate labels for all samples, the same algorithm is applied to consolidate the labels, resulting in a refined set of malware class-family labels. By doing so, it reduces the time needed for reverse engineering and consulting resources like Malpedia \cite{Malpedia} and other open databases to confirm assigned labels. Analyzing samples randomly would significantly prolong the classification process, as it would require a repeated search for related information. By grouping samples that predominantly belong to the same category, dataset construction becomes more efficient, minimizing the need for redundant searches.

Additionally, this paper analyzed the malware programs using dynamic analysis with a prominent open-source malware analysis system, Cuckoo Sandbox \cite{cuckoo}, for five minutes, and performed manual analysis, such as reverse engineering, on each sample. We employed Cohen’s Kappa coefficient \cite{vieira2010cohen} to assess the agreement between the two authors during the labeling process. Initially, they independently labeled 5 percent of the malware programs, resulting in a Cohen’s Kappa coefficient of approximately 0.78. Following a training discussion, they labeled an additional 10 percent of the programs, including the initial 5 percent, achieving a significantly improved Cohen’s Kappa coefficient of 0.94. After resolving discrepancies in differently labeled programs, the authors proceeded to label the remaining programs in nine iterative rounds, each including an additional 10 percent. Throughout this iterative process, Cohen’s Kappa coefficient consistently remained above 0.9. In every iteration, any disagreements were addressed and resolved collaboratively with a third author. By the end of this process, all malware programs were labeled consistently. We then reached a consensus on the classification of each sample, grouping them into their respective categories by combining classes and families, thereby finalizing the creation of our dataset.

\subsubsection{Binary Analysis Framework} radare2\cite{eom2024r2i}is a widely-used multi-architecture, cross-platform binary analysis toolkit known for its capability to perform various static analysis and manage different type of program representations on binaries. We use radare2 as our binary analysis framework to perform disassembly on malware binaries, as other frameworks, such as angr, often overlook system functions. This process involves generating instructions, partitioning them into basic blocks and functions, and extracting key program metadata, including both opcode and operand decoding. For instance, the binary sequence ``B8 04 00 00 00'' is disassembled into the assembly instruction ``mov eax, 4'' based on the x86 instruction set, where B8'' specifies the opcode for moving an immediate value into the ``eax'' register, and ``04 00 00 00'' represents the 32-bit immediate value 4. After recovering the instructions, radare2 identifies a collection of instructions with a single entry point and a single exit point as basic blocks. A basic block's entry point is typically the first instruction in a function, the instruction immediately following a branch instruction, or the target of a branch instruction, such as jump or call. Conversely, branch instructions serve as the exit point of a basic block.

While generating function metadata, we observed that some function names are prefixed with ``fcn.'' followed by their function addresses. It is important to note that all malware programs analyzed are stripped binaries, meaning their symbol tables have been removed. As a result, the original user function names cannot be directly retrieved.  Instead, we represent these functions by combining ``fcn.'' with their respective addresses. For the remaining functions, they typically correspond to Windows system functions, which are identified by names starting with ``sym.imp.'' Accordingly, we collect basic block metadata and instruction metadata from both user-defined functions and system functions. This program metadata includes metrics such as the average instruction length ($avg\_instr$), average basic block size ($avg\_bb$), etc., all measured in bytes. These metrics are subsequently employed by SemanticForge for further analysis.

\subsubsection{SemanticForge}
Malware variants belonging to the same class often reuse code \cite{Calleja}. Utilizing functions as the fundamental unit for malware classification poses challenges to accuracy. This is because functions can replace a function call site with the body of the called function due to compiler or manual optimizations. Furthermore, accurately identifying function boundaries is inherently difficult \cite{Xiaozhu,Andriesse}. Binary analysis frameworks like radare2, angr, IDA, etc., also struggle to precisely recover function boundaries in binary programs, rendering the use of functions as a detection unit impractical. Besides, employing instructions as the basic unit also presents challenges as individual instructions lack sufficient semantics.

Therefore, we choose basic blocks as the foundational units for our design. A basic block is a core semantic and control unit within a program, consisting of a sequence of instructions that execute sequentially. Once the first instruction in the sequence is executed, all subsequent instructions are guaranteed to execute in order, with no external instructions intervening until the final instruction in the block is completed. In essence, a basic block is a linear segment of code that contains no branches, except at its entry and exit points. Leveraging basic blocks as the basis for malware classification streamlines the analysis process by breaking down complex programs into smaller, more manageable units. Analyzing basic blocks individually enhances understanding of program behavior, as each block typically encapsulates a specific set of instructions corresponding to a particular functionality. Furthermore, basic blocks act as natural boundaries for various analysis techniques, providing discrete units of code that can be executed atomically. This allows analysis algorithms to focus on specific sections of a program rather than the entire codebase, enabling more efficient and scalable analysis.

In this paper, we introduce a novel local and global self-analysis technique to capture the local similarity within basic blocks and global correlations of basic blocks while maintaining the similarity of basic blocks among malware programs from the same category and preserving correlations between basic blocks in the same malware sample. Local similarity, in this context, is closely associated with the statistical co-occurrence of instruction sequences observed across programs, quantified by Mutual Information (MI). Traditionally, the  joint statistics of instructions are calculated from individual programs before being compared across different ones. However, the majority of current approaches are restricted to analyzing the joint occurrence patterns of instruction-level metrics, such as byte codes, opcodes, operands, or assembly characters, and struggle to extend to larger, more meaningful patterns like basic blocks due to the curse of dimensionality, particularly in the case of MI. Additionally, these methods typically assume global statistical co-occurrence within the entire program, which is often an overly strong assumption. In contrast, our approach focuses on measuring similarity locally within a surrounding basic block, rather than globally across the entire program. To account for the mutual relationships between local basic blocks within a program, we apply a CLAHE algorithm \cite{clahe} to enhance the correlations between them. Therefore, our framework explicitly accounts for both local and global information. Next, we describe the local similarity calculation within basic blocks and mutual relationship enhancement between basic blocks to obtain self-analysis descriptors.

\textit{5.1) Local Similarity Calculation:} Our objective is to compare one malware $T(x, y)$ with another $M(x, y)$, acknowledging that $T$ and $M$ may differ in dimensions. While assessing similarity between objects can be challenging, evaluating similarity within each malware variant is more straightforward. This evaluation can be performed using simple measures, such as the sum of squared distances, which reveal local similarities that can then be compared across different malware programs.  However, the fixed calling conventions of x86 or x64 introduce a challenge: comparing instructions within the calling convention to those in a basic block can result in similar sums of squared distances for equivalent instructions from different basic blocks that terminate with call sites. This similarity complicates the differentiation of malware programs. To address this, we opt to use the centered instruction for comparison. Additionally, the varying sizes of basic blocks make it impractical to encode them using a fixed number of local similarity descriptors. Individually analyzing each basic block is computationally expensive, often doubling processing time, as it involves dividing large basic blocks into smaller chunks and identifying their center instructions. This process may not fully exploit the advantages of static analysis. Additionally, due to the varying lengths of instructions, direct comparison is infeasible. To overcome these challenges, we account for the magnitude of each basic block by using the average instruction length in conjunction with the surrounding basic block's average size. The mean serves as a representative value, capturing the dataset's typical characteristics while encompassing the influence of outliers  for special basic blocks relevant to our analysis. To mitigate the impact of instruction disorder caused by using average instruction length and basic block size, we compute local similarity descriptors densely across the surrounding basic block. We then employ a binned log-polar representation \cite{belongie2002shape} to address the increasing positional uncertainty associated with distance.

We assign a local similarity descriptor $lsd_{x}$ to each instruction $x$ by correlating it with a surrounding basic block and generating a local correlation matrix. Here, ``local'' refers to a tiny share of the malware, such as 1\%, as opposed to analyzing the entire malware. The correlation matrix quantitatively measures the similarity between instructions, with higher values reflecting greater similarity. By incorporating the exponential function (see Equation \ref{correlation}), the matrix amplifies small differences between instructions, resulting in relatively large similarity values. This approach enhances robustness to minor variations while imposing greater penalties for significant discrepancies. The correlation matrix is then densely calculated and  transformed into a binned log-polar format. This format offers two key advantages: (i) It yields a informative descriptor $lsd_{x}$ for each instruction $x$. (ii) It accommodates growing positional uncertainty as the distance from instruction $x$ increases, thereby addressing its local spatial affine deformations. To match an entire malware $T$ to $M$, we densely compute the local similarity descriptors $ld_{x}$ throughout $T$ and $M$ for every other basic block. It's worth noting that this approach significantly expedites the computation of local similarity descriptors because each basic block's processing is independent. To optimize efficiency, we parallelize the computation using multi-threading.
The local similarity descriptors in $T$ are combined to create a unified global ensemble, preserving their respective geometric relationships. For $T$ to successfully match within $M$, a comparable ensemble of local similarity descriptors must be identified in $T$ – ensuring alignment in both descriptor values and their geometric layouts.

This paragraph introduces the process for calculating the local similarity descriptor $lsd_{x}$ associated with an instruction $x$. $x$ is compared to other instructions within a larger basic block centered around $x$, using the sum of squared distance (SSD) of their bytecode. Each local similarity point in $lsd_{x}$ is computed according to Equation \eqref{SSD}, where $i$ represents each byte code in the instruction and $y$ denotes an instruction different from $x$ in the surrounding basic block. The distance matrix $SSD(x, Y)$ is transformed into a ``correlation matrix'' $CM(x, Y)$ as per Equation \eqref{correlation}. The correlation matrix $CM(x, Y)$ is then converted into log-polar coordinates, centered at $x$, and partitioned into $m*n$ bins where $m$ represents angular divisions and $n$ denotes radial intervals. The maximal values typically correspond to instructions exhibiting the most similar semantics, highlighting the key features of the basic block. The highest value from each bin is selected to create  the $m*n$ components of our local similarity descriptor $lsd_{x}$ for instruction $x$. Consequently, our descriptor mitigates additional local non-rigid deformations. Finally, semanticForge will produce a global ensemble of local similarity descriptors, which are subsequently normalized by linearly adjusting their values to fall within the range [0..1] using Equation \eqref{normalization} and Equation \eqref{scale} to ensure invariance to differences in the instruction selection and convert them to a feature map. Each  feature map is reshaped to a size $(w, h, m*n)$ following various recommended fixed feature map heights with variable widths based on the size of global local similarity descriptors as specified in \cite{Nataraj}. These feature maps are further enhanced by a mutual relationship enhancement method.

\begin{equation}\label{SSD}
SSD(x, Y) = \{ \sum_{i=0}^l (x_{i} - y_{i})^2 \mid y \in Y \}
\end{equation}

\begin{equation}\label{correlation}
	CM\left ( x,Y \right )=exp^{-SSD\left ( x,Y \right )}
\end{equation}

\begin{equation}\label{normalization}
CM_{std} = \frac{CM(X, Y) - \min(CM(X, Y))}{\max(CM(X, Y)) - \min(CM(X, Y))}
\end{equation}

\begin{align}\label{scale}
CM_{scl} &= CM_{std} \cdot \big(\max(CM(X, Y)) - \min(CM(X, Y))\big) \notag \\
&\quad + \min(CM(X, Y))
\end{align}

In conclusion, the characteristics and benefits of the local similarity descriptor can be summarized as follows: i) local similarity descriptors are treated as intrinsic characteristics of basic blocks, hence they are evaluated locally within the scope of a neighboring basic block, rather than globally across the entire malware. This methodology widens the utility of the descriptor, making it applicable to a diverse array of complex malware variants. ii) The log-polar format integrates incorporates local deformations caused by inconsistencies in instruction alignment into the descriptor. Selecting the maximum value from each bin reduces the descriptor's sensitivity to the exact location of the optimal match within the bin.
iii) As the bins dynamically adjust in size with alterations in the future malware distribution, this capability supports the accommodation of additional radially expanding non-rigid changes over time. iv) Using basic blocks as the basic unit for evaluating internal features enables the identification of more significant functional patterns compared to examining individual instructions independently.

\begin{figure}[!htb]
	 	\centering
	 	\includegraphics[scale=0.35]{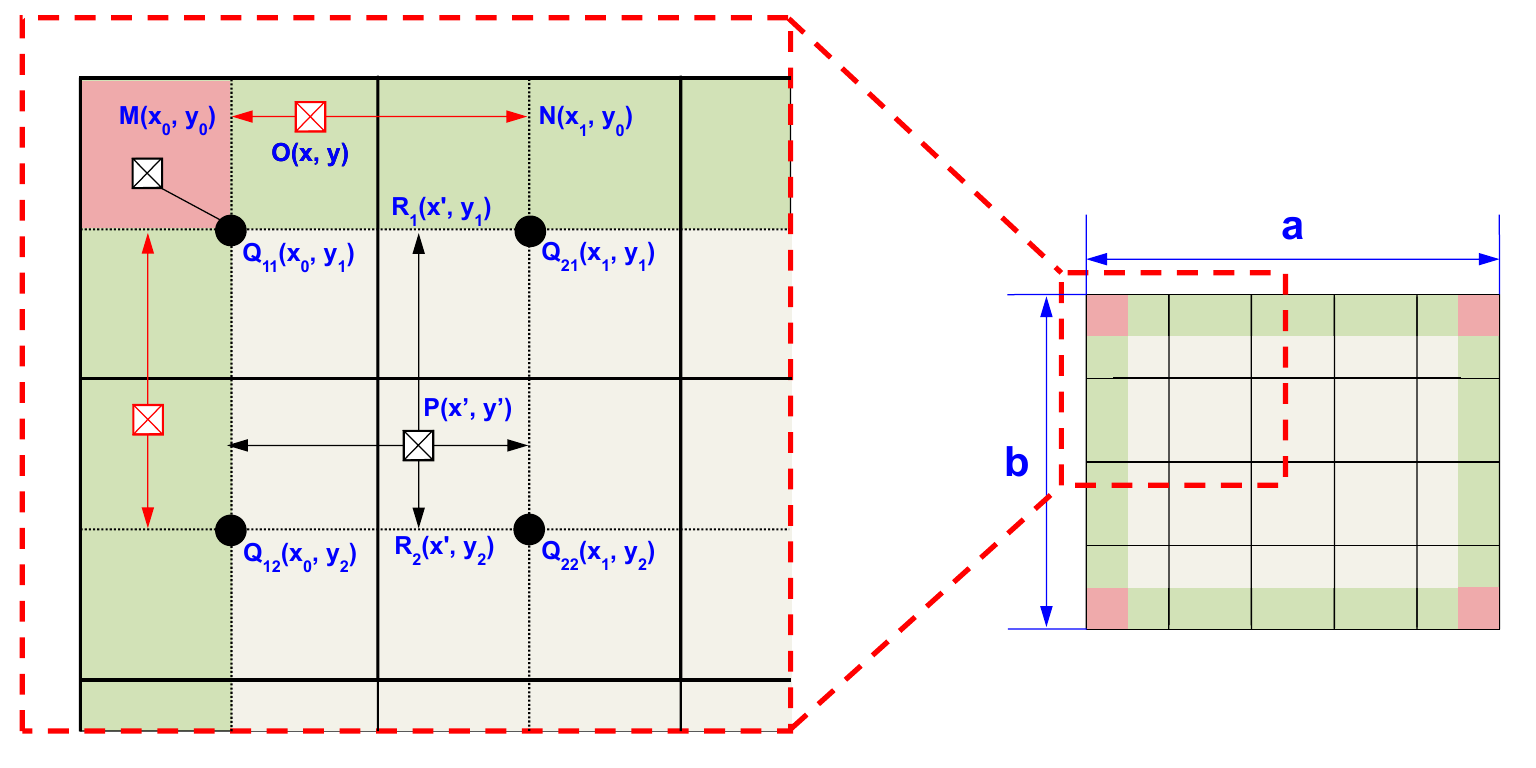}
	 	\caption{Transformation of Input Descriptors}
	 	\label{fig:transformation}
	\end{figure}

\textit{5.2) Mutual Relationship Enhancement} This step takes as input a feature map from the last step. Our aim is to identify similar collections of  descriptors across different malware programs. Additionally, these local similarity descriptors may be located at various positions within feature maps. To enhance the classifier's accuracy, it's crucial to capture these similar patterns within feature maps. In this paper, a feature map is essentially an image. Motivated by the observation that the useful information in an image is typically characterized by closely contrasted pixels \cite{equalization}, the image intensities are adjusted to improve contrast using a CLAHE algorithm and meanwhile enhance the correlation between neighboring local similarity descriptors. Since the influence of the similarity or dissimilarity between neighboring descriptors in terms of intensity is not significant for our study, the transformation of image intensities does not affect the classification process. This adjustment enables better distribution of intensities across the histogram for a local region, thereby amplifying the similarity across different malware programs. Consequently, this enhancement increases the likelihood of identifying similar patterns within the feature maps. Metaphorically, CLAHE can be likened to a signal transformation observed in biological neural networks, where the aim is to maximize the firing rate of neurons based on input statistics \cite{enhancement}. This process involves four phases: Division, Histogram, Clipping, and Transformation. In the Division phase, the feature map is segmented into smaller regions, usually represented by a grid of dimensions $(a \cdot b)$, where $a$ denotes the number of divisions along the width of the map and $b$ represents the number of divisions along its height. The Histogram phase involves constructing a cumulative distribution function that reflects the frequency of descriptor values across the region. This function helps in understanding the distribution of descriptor intensities within each region. The calculation of the histogram is depicted in Equation \eqref{histogram}, where $P$ represents the total number of pixels values in a region, typically 256, and $n_x$ denotes the total count of a particular descriptor intensity within a region. In the clipping phase, the objective is to redistribute descriptor evenly to prevent the amplification of noise within a region. This redistribution occurs when the difference in intensity between neighboring descriptors exceeds a predefined threshold $cl$ across the entire histogram. In the histogram, each descriptor value corresponds to a bin, and the value of each bin represents a cumulative count obtained by summing the frequencies from descriptor intensity 0 up to a specific descriptor value. For example, the value of the $i_{\text{th}}$ bin at the region $x$, denoted as $cdf_x(i)$, is precisely computed using Equation \eqref{histogram} by summing the frequencies from descriptor intensity 0 to descriptor intensity $i$. Next, the slope $k$ between two neighboring descriptor intensities can be computed using Equation \eqref{slope} where $cdf_x{(i)}$ and $cdf_x{(i-1)}$ represent the values of two neighboring descriptor intensities and $i-(i-1)$ equals to 1. The redistribution of a high-frequency descriptor intensity can decrease the slope for its neighboring descriptor intensity, thereby limiting the amplification of such descriptors with high frequency. Noises typically exhibit similar descriptor values, leading to regions with homogeneous descriptor values. In such regions, the histogram is highly peaked around these descriptor intensities \cite{clahe}. We utilize clipping to alleviate the contrast amplification for noises induced by 0 paddings in malware programs. Consequently, the histogram is adjusted to a newly generated cumulative distribution. In the transformation phase, various transformations are applied to map the input descriptors to output descriptors across the entire region based on their positions (see Fig. \ref{fig:transformation}). As detailed in \cite{ahe}, descriptors near corners shaded pink and the center descriptors (black points) in each region are transformed using the function defined by Equation \eqref{transfunc}, where ${cdf_{\min}}$ represents the smallest non-zero value of the cumulative distribution function and ${cdf_{\max}}$ represents the maximum value in the cumulative distribution function. This transformation function may expand the values in a map region to a wider range.
\begin{equation}\label{histogram}
    cd{f_x}(i) = \left\{ \sum_{j=0}^i n_x(x = j) \,\middle|\, 0 \leq i < P \right\}
\end{equation}

	 \begin{equation}\label{slope}
	     k = \frac{{{cdf_x{(i)}} - cdf_x{(i-1)}}}{{{i} - {(i-1)}}}
	 \end{equation}
 
\begin{equation}\label{transfunc}
    h(x) = \operatorname{round}\left(\frac{\operatorname{cdf}(x) - cdf_{\min}}{cdf_{\max} - cdf_{\min}} \cdot (L - 1)\right)
\end{equation}

	Descriptors near the boundary shaded light green undergo linear interpolation. The transformation of each input descriptors can be calculated using \eqref{eq:linear}. Additionally, descriptors in the central region of the map shaded light grey undergo bi-linear interpolation. This means that the input descriptor (black square) is transformed into the output descriptor based on the four neighboring cumulative distribution functions, following the equations shown in \eqref{eq:R1}, \eqref{eq:R2}, and \eqref{eq:P}.

	\begin{equation}\label{eq:linear}
	 h(O) = \frac{{{x_1} - x}}{{{x_1} - {x_0}}}h({M}) + \frac{{x - {x_0}}}{{{x_1} - {x_0}}}h({N})
	\end{equation}
	
	\begin{equation}\label{eq:R1}
	 	h({R_1}) = \frac{{{x_1} - x'}}{{{x_1} - {x_0}}}h({Q_{11}}) + \frac{{x' - {x_0}}}{{{x_1} - {x_0}}}h({Q_{21}})
	\end{equation}
	 
	\begin{equation}\label{eq:R2}
      h({R_2}) = \frac{{{x_1} - x'}}{{{x_1} - {x_0}}}h({Q_{12}}) + \frac{{x' - {x_0}}}{{{x_1} - {x_0}}}h({Q_{22}})
	\end{equation}
	 
	\begin{equation}\label{eq:P}
	 {{h}}(P) = \frac{{{y_2} - y'}}{{{y_2} - {y_1}}}h({R_1}) + \frac{{y' - {y_1}}}{{{y_2} - {y_1}}}h({R_2})
	\end{equation}
Following the transformation, the resulting feature map is resized to have dimensions $(w1, h1,m*n)$, which serves as input to the classifier.

\section{Evaluation Settings}
\label{sec:setup}
\subsection{Evaluation Metrics}
The performance metrics of \textit{VisUnpack} include \emph{accuracy}, \emph{computational efficiency} and \emph{resource consumption}. \emph{accuracy} can be evaluated using commonly used machine learning parameters: Precision, Recall, F1, and Accuracy. \emph{computational efficiency} can be assessed by the processing time per malware program, denoted by \emph{$\overline{T}$}. \emph{resource consumption} can be evaluated by the size of the dataset used by the machine learning model. Let \emph{T} represent the total clock ticks used to process a certain number of malware programs, denoted by $num\_files$. We can compute \emph{avg\_time} as:

\begin{equation}\label{eq:efficiency}
	\overline{T} = \frac{T}{num\_files}.
\end{equation}

\begin{table}
	\footnotesize
	\centering
	\caption{VisUnpack Malware Dataset}
	\begin{tabular}{cc|cc}
		\hline
		Family.Subfamily&No. of Files & Family.Subfamily&No. of Files\\
		\hline
		adware.bundler & 334/0 &adware.domaiq  & 465/122\\
		
		adware.ibryte & 151/0&adware.imali  & 1209/1209\\
		
		adware.loadmoney & 201/0&adware.multiplug  & 3029/6\\
		
		adware.softpulse & 217/115&worm.allaple  & 561/2\\
		
		 trojan.antifw &973/361&trojan.autorun  & 90/0 \\
		
		worm.autorun  &92/0 & trojan.dridex&4995/786 \\
		
		trojan.ekstak  & 152/0&trojan.emotet  & 3082/0\\
		
		virus.expiro  &1927/0&trojan.fareit  & 362/362 \\
		
		trojan.gandcrab & 150/0&trojan.lokibot  & 399/0\\
		
		trojan.morstar  & 126/0&worm.mydoom & 109/109\\
		
		pua.playtech  & 116/116&pua.syncopate  & 171/171\\
		
		pua.toolbar  & 437/437&trojan.python  & 137/3\\
		
		trojan.servstart  & 117/0&trojan.startpage  & 402/280\\
		
		trojan.ash  & 137/137&trojan.cobaltstrike  & 98/98\\
		
		trojan.dacic  & 124/124&trojan.hijacker  & 125/125\\
		
		trojan.jqxm  & 526/0&trojan.qakbot  & 95/95\\
		trojan.rozena  & 98/98&trojan.softpulse  & 104/104\\
		trojan.swisyn  & 442/442&trojan.toolbar  & 389/389\\
		
		trojan.tzl  & 850/0&trojan.unruy  & 103/0\\
		
		trojan.wacapew  & 289/289&trojan.wannacry  & 1394/565\\
		
		trojan.webcake  & 831/831&trojan.xpack  & 795/795\\
        trojan.zusy  & 101/101&virus.virlock & 230/0\\
        virus.sality  & 228/0&worm.winhash  & 143/0\\		
		\hline
	\end{tabular}
	\label{tab:dataset}
\end{table}

\subsection{Experiment Configurations}
\emph{Equipment and Dataset}. The \textit{VisUnpack} system operates on a standard PC featuring 32 logical processors and 24 cores, with 64.0GB of available RAM, running the 64-bit Microsoft Windows 11 Home operating system. Each processor uses a 13th Gen Intel(R) Core(TM) i9-14900k CPU operating at 3.2GHz. The development environment utilizes PyTorch version 2.5.1, Pillow version 11.0.0, and radare2 version 1.9.2. Malware programs were sourced from VirusShare \cite{virusshare} across eight years from 2017 to 2024. We filtered out malware programs that are not Windows PE and do not have complete program structures. Therefore, our dataset reflects the trend of Windows malware in recent years. It consists of a total of 27,106 malware programs, distributed across 46 malware categories, with a diverse distribution of malware per category, as shown in TABLE \ref{tab:dataset}.The numbers on the left side of the `/' represent the total number of malware programs, while those on the right indicate the number of packed programs. Adware and trojans stand out as dominent types of malware in the wild due to their profit-driven nature. Adware generates revenue by displaying advertisements, while trojans serve various malicious purposes, such as stealing sensitive information or enabling other cybercrimes. The potential financial gains motivate attackers to produce and disseminate these types of malware on a larger scale. PUAs have also caught our attention. These applications are often bundled with legitimate software, and users may inadvertently install PUAs while installing the primary software they intended to download.

\emph{Parameters.} In order to enhance our detection capabilities and minimize the risk of false negatives, we have deliberately set the number of considered substrings $t$ to 3. This strategic decision is based on empirical evidence showing that malware programs combining functionality from more than three families are often classified as new variants \cite{classguide}. We set the threshold $thr$ to 0.75 derived from statistical information on substring similarity. Additionally, to streamline semanticForge's operations without compromising on efficacy, we consider the average instruction length, which we have empirically determined to be 2.97 bytes. Consequently, we use three bytes as the comparison unit, ensuring a granular analysis while minimizing computational overhead. Similarly, in pursuit of computational efficiency, we  employ a comparison region of 15 bytes in terms of $avg\_bb$. To generate  feature maps, $m$ and $n$ are set to 1 and 3, respectively. $a$ and $b$ are both set to 8, taking into account the trade-off between local and global contrast enhancement. Smaller tile sizes enable more localized contrast enhancement, which is beneficial for capturing fine details in the feature maps. However, excessively small tile sizes may lead to over-amplification of noise. By selecting such a tile size, we aim to strike a balance between local and global enhancement. To balance enhancement and noise suppression further, we set the $cl$ to 4. This value is commonly chosen based on empirical observations and practical considerations \cite{ahe}. Higher clip limits can result in more aggressive contrast enhancement but may also amplify noise, while lower clip limits may provide less enhancement but can suppress noise more effectively.

In identifying a suitable classifier structure, we conducted experiments with various configurations of the popular VGG network, ranging from VGG11 to VGG19. Through this exploration, we found that VGG11 achieved the optimal balance between accuracy and efficiency for our task. The constructed classifier consists of five convolutional blocks and one fully connected block, with detailed model parameters provided in \cite{simonyan2014very}. Since the VGG network requires a fixed input size, we analyzed the distribution of shapes for feature maps to determine suitable dimensions. We observed that a majority of  feature maps could be accommodated when the width and height reached 512 and 128, respectively. 
The datasets used in this study exhibit highly unbalanced class distributions. To ensure that batches contain representative samples from all classes, we implemented a sampling strategy favoring minority categorizations. Specifically, we assigned a higher probability of selection to samples from minority categorizations, making the probability inversely proportional to the ratio of the number of samples in a categorization to the total number of samples. This approach ensures that minority class samples are more likely to be included in each batch, while still allowing for representation from majority classes. For our training process, we allocated 90\% of the samples randomly from each class for training, reserving the remaining 10\% for testing purposes. Our VGG11 model is trained for 100 epochs using stochastic gradient descent. Then, malware programs in the test dataset are used to check VGG11’s generalization ability.

\section{Evaluation Results}
    \label{sec:evaluation}

\subsection{Accuracy}

	  We will compare VisUnpack with two different types of data visualization based classification methods: (1) advanced image processing, including DRBA \cite{Zhihua}, GIST \cite{Nataraj}, VisMal \cite{Malbrain}, and \textit{VisUnpack\_fe}; and (2) advanced feature extraction, such as DNN \cite{Saxe}, KNN \cite{Ouahab}, CNN \cite{Xusheng},Autoencoder \cite{Xiang}, CNN\_attention \cite{Yakura}, MDMC \cite{Baoguo}, LBP \cite{liu2019new}, Chen \cite{chen2020malware}, Adkins \cite{adkins2013heuristic}, \textit{VisUnpack\_grey} and \textit{VisUnpack\_ssd}. Due to the unavailability of source codes, we have re-implemented the aforementioned algorithms for classification. DRBA \cite{Zhihua} employs an image augmentation technique to improve image quality by enhancing rotation range, width shift, height shift, and other properties. GIST \cite{Nataraj} analyzes the statistical distribution of pixels in malware images using a two-dimensional Gabor function, adjusted for specific frequencies and orientations via a Gaussian envelope. By modifying these parameters, it generates Gabor filters to extract features across eight orientations and four scales. VisMal \cite{Malbrain} emphasizes the geometric arrangement of malware pixels derived from bytecode representations. It uses a modified histogram equalization algorithm to strengthen pixel correlations, focusing on local neighborhood relationships. Deep Neural Network (DNN) \cite{Saxe} captures the linear relationships within malware bytecode, learning feature embeddings for classification. In contrast, the K-Nearest Neighbors (KNN) method \cite{Ouahab}, a non-parametric approach, predicts classifications based on data point similarity through distance comparisons. Approaches like Convolutional Neural Networks (CNN) \cite{Xusheng}, autoencoders \cite{Xiang}, and CNN with attention mechanisms (CNN\_attention) \cite{Yakura} focus on learning both local and global correlations within bytecode, leveraging these relationships for classification. Mutual Dependency Markov Chains (MDMC) \cite{Baoguo} highlights the Mutual Information among neighboring bytecode and calculate its joint probabilistic
 matrix as features for classification. LBP \cite{liu2019new} uses dense SIFT to extract local features from various locations and scales in images. A subset of these features is grouped into clusters via k-means clustering, with the cluster centers forming a bag-of-visual-words representation, encoding both local and global feature information. Chen \emph{et al.} \cite{chen2020malware} extracted opcodes from basic blocks of malware, processed them using the SimHash algorithm to generate fingerprints, and transformed these fingerprints into pixel values to create images by concatenating pixel data. Adkins \emph{et al.} \cite{adkins2013heuristic} replaced registers, memory locations, constants, and variables with standardized labels such as `REG’, `LOC’, `MEM’, `CONST’, and `VAR’. Four instructions were grouped to form n-grams, which were processed with MD5. The truncated bits of the MD5 sum were used as a feature hash index. In contrast, our \textit{VisUnpack} approach goes beyond local connections by incorporating internal semantics and local similarity within basic blocks while capturing correlations between them at both local and global levels. Shallow layers of VGG11 focus on learning local connections, while deeper layers identify and exploit global relationships for more robust classification. Next, we evaluate the accuracy of \textit{VisUnpack} on both packed and unpacked versions of the Malfiner datasets before comparing it with other methods employing diverse techniques.

\textit{1) Performance of VisUnpack} As shown in Table \ref{tab:accuracy1}, \textit{VisUnpack} achieves outstanding average accuracy, precision, recall, and F1 score metrics on the unpacked Malfiner dataset, each standing at 99.7\%. Among the 46 examined malware families, 89\% achieve perfect classification accuracy of 100\%. Fig. \ref{fig:similarity} highlights that malware programs within the same category exhibit similar textures, reflecting shared code bases, whereas different categories display distinct textures and indicate unique code bases. It is worth noting that malware programs utilizing anti-emulation, anti-debugging, code obfuscation, multi-threaded, or polymorphic techniques, such as worm.allaple, and virus.virlock, can also be detected with perfect accuracy at 100\%. Besides, malware programs belonging to the same family with different class can be detected with very high accuracy, such as adware.softpulse (100\%), pua.softpulse (100\%), pua.toolbar (100\%), trojan.toobar (100\%), trojan.autorun (100\%) and worm.autorun (86\%). 
\begin{table}
		\centering
		\caption{Accuracy (\%) on Packed/Unpacked Malfiner}
		\begin{tabular}{|p{2cm}|p{1.15cm}|p{1.15cm}|p{1.15cm}|p{1.15cm}|}
			\hline
			& Precision &Recall& F1&Accuracy\\
			\hline
			 DNN \cite{Saxe} & 0.01/95.3 & 1/94.2& 0.02/94.5&1/94.2\\
			\hline
			KNN \cite{Ouahab}& 84.8/95 & 82/92.8&81.9/92.7&82/92.8\\
			\hline
            CNN \cite{Xusheng}& 0.01/0.01 & 1/1&0.02/0.02&1/1\\
			\hline
            Autoencoder \cite{Xiang}& 0.001/0.02 & 0.4/1.5&0.004/0.05&0.4/1.5\\
			\hline 
            CNN\_attention \cite{Yakura}& 0.07/1.5 & 2.7/12.2&0.1/2.7&2.7/12.2\\
			\hline
            DRBA \cite{Zhihua}& 90.2/99.3 & 89/99.3&87.5/99.3&89/99.3\\
			\hline  
            GIST \cite{Nataraj} & 89.4/98.8 &89.1/98.7 &89.2/98.7&89.1/98.7 \\
            \hline
            VisMal \cite{Malbrain} & 94.3/99.7 &93/99.7 &92.1/99.7 &93/99.7\\
            \hline
            MDMC \cite{Baoguo}& 92.4/99.4 & 92.2/99.4&91.4/99.4&92.2/99.4\\
            \hline
            VirusTotal \cite{virustotal} & - &- &-&14.2\\
            \hline
            LBP \cite{liu2019new} & 90.8/98.9 &90.4/98.9 &90.4/98.9&90.4/98.9\\
            \hline
            Chen \emph{et al.} \cite{chen2020malware} & 89.5/97.8 &89.9/97.7&88.4/97.6 &89.9/97.7\\
            \hline
            Adkins \emph{et al.} \cite{adkins2013heuristic} & 87.9/98.9 &87.3/99 &86.1/98.7&87.3/99\\
            \hline
            VisUnpack\_grey & 93.7/99.7 & 93.1/99.7 & 92/99.7 &93.1/99.7\\
            \hline
            VisUnpack\_ssd & 96/99.6 &95.6/99.6&95.3/99.6 &95.6/99.6\\
            \hline
            VisUnpack\_fe & 92.7/99.7 &93.2/99.7 &92/99.7 &93.2/99.7\\
            \hline
            VisUnpack& 96.6/99.7 &96.1/99.7&95.9/99.7 &96.1/99.7\\
            \hline
		\end{tabular}
		\label{tab:accuracy1}
	\end{table}

  

\begin{figure}[!htb]
    \centering
    \label{fig:similarity}
    \includegraphics[width=2.8cm,height=2cm]{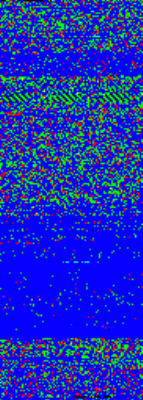}
    \includegraphics[width=2.8cm,height=2cm]{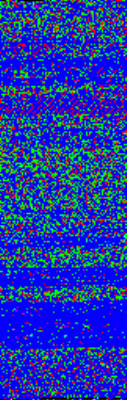}
    
    \vspace{0.5em}

    \includegraphics[width=2.8cm,height=2cm]{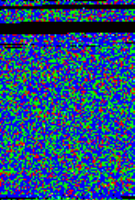}
    \includegraphics[width=2.8cm,height=2cm]{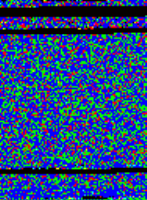}

    \vspace{0.5em}
      \includegraphics[width=2.8cm,height=2cm]{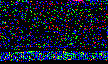} \includegraphics[width=2.8cm,height=2cm]{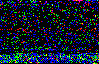} 
    \caption{Malware Similarity in virus.virlock, trojan.dridex, and trojan.lokibot}
     \label{fig:similarity}
\end{figure}

\begin{figure}[!htb]
  
    \centering
    
    \includegraphics[width=2.8cm,height=2cm]{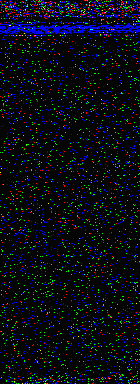}
    \includegraphics[width=2.8cm,height=2cm]{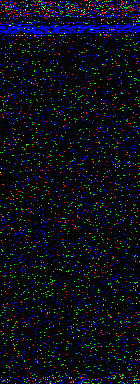}
    
    \vspace{0.5em}

    \includegraphics[width=2.8cm,height=2cm]{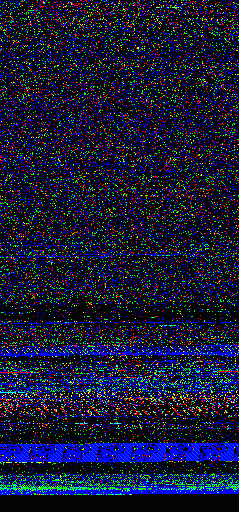}
    \includegraphics[width=2.8cm,height=2cm]{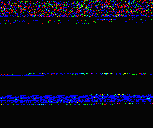}

    \caption{packed/unapcked pua.playtech sample (left) and packed/unpacked trojan.xpack sample (right).}
    \centering
      \label{fig:playtech}
\end{figure}

When samples are packed on Malfiner dataset, the average accuracy, precision, recall, and F1 score drop to 96.6\%, 96.1\%, 95.9\%, and 96.1\%, respectively, representing decreases of 3.2\%, 3.7\%, 4.0\%, and 3.2\%. This reduction is particularly significant for categories like \textit{pua.playtech}, \textit{trojan.swisyn}, \textit{trojan.toolbar}, \textit{trojan.wacapew}, and \textit{trojan.xpack}, with accuracies dropping to 75.0\%, 23.3\%, 70.7\%, 13.0\%, and 76.4\%, respectively. Specifically, \textit{pua.playtech} and \textit{trojan.xpack} are frequently misclassified as each other due to their use of the same NSIS packer, which generates highly similar code bases. After unpacking, \textit{VisUnpack} achieves perfect classification by leveraging the distinct textures of the unpacked programs, as shown in Fig. \ref{fig:playtech}. Similarly, malware from \textit{trojan.swisyn} and \textit{trojan.wacapew} are misclassified as \textit{trojan.fareit} and \textit{trojan.toolbar}, respectively, because these programs are packed together into a single program, making accurate classification nearly impossible without unpacking. Some malware categories, such as \textit{adware.domaiq}, \textit{adware.imali}, \textit{trojan.dridex}, and \textit{trojan.wannacry}, achieve 98\% or above accuracy even when packed by various packers, such as UPX\cite{upx}, PECompact\cite{PECompact}, Petite\cite{petite}, and
MPRESS\cite{mpress}, demonstrating that certain packers influence classification more than others. Besides, if multiple types of malware are packed together, accurate classification becomes challenging, underscoring the importance of unpacking for reliable results. This finding challenges the assumption in prior works \cite{Saxe,Ouahab, Nataraj, Yakura, Malbrain, liu2019new} that packing has minimal impact on malware classification. While some studies, such as \cite{chen2020malware} and \cite{adkins2013heuristic}, explicitly use unpacked datasets, most others fail to address the reliability of their methods when applied to packed samples.




\textit{2) Comparison between Different Methods} When analyzing other classification methods on the packed and unpacked Malfiner dataset, as shown in Table \ref{tab:accuracy1}, nearly all methods exhibit significant improvements in accuracy when the programs are unpacked compared to their packed counterparts. This further highlights the detrimental impact of packing on malware classification and underscores the necessity of unpacking programs before developing reliable classification techniques. Notably, DNN \cite{Saxe} achieves only 1\% accuracy on malware gray images derived from the packed Malfiner dataset, compared to 94.2\% on the unpacked dataset. This stark difference indicates that packing disrupts the linear relationships among malware bytecodes that DNNs rely on for classification. KNN \cite{Ouahab}, while achieving a relatively high accuracy of 89\% on packed malware gray images (rising to 92.8\% on unpacked images), performs poorly on 34.8\% of malware categories, with accuracies below 50\%. This suggests that distance metrics may not effectively distinguish malware belonging to the same class and family. Furthermore, CNN \cite{Xusheng}, Autoencoder \cite{Xiang}, and CNN\_attention \cite{Yakura} perform extremely poorly, with accuracies below 12.2\% even on the unpacked Malfiner dataset. This indicates that local features with simple abstraction, reconstruction error, and the higher importance of certain bytes are not enough to capture meaningful bytecode patterns or correlations required for accurate classification. In contrast, VisUnpack\_grey, which applies VGG11 directly to malware gray images from packed and unpacked Malfiner datasets, achieves accuracies of 93.7\% and 99.7\%, respectively. This demonstrates that VGG11 is significantly more effective at capturing similarities for malware classification compared to aforementioned neural network architectures, making it a strong foundation for the \textit{VisUnpack} framework.

GIST \cite{Nataraj} performs much better than other methods, with an accuracy of 89.1\% and 98.7\% on packed and unpacked malware grey images, suggesting that malware programs within the same class and family often have similar statistical features, unless they change significantly. This aligns with the idea of malware variants. DRBA \cite{Zhihua}, MDMC \cite{Baoguo} and VisMal \cite{Malbrain} have a higher accuracy than GIST on unpacked malware grey images, with an accuracy of 99.3\%, 99.4\%, and 99.7\% respectively. However, DRBA and MDMC show lower accuracy than VisMal on both packed and unpacked images, indicating that methods focusing on enhancing local neighborhood relationships are less sensitive to packing compared to those emphasizing global neighborhood relationships. Surprisingly, LBP, which considers both local and global features, achieves accuracies of only 90.4\% and 98.9\% on packed and unpacked malware gray images, respectively, performing worse than methods relying solely on either local or global features. Moreover, the anti-virus products in VirusTotal exhibit limited effectiveness, with a modest average accuracy of 14.2\%, highlighting their inability to accurately identify malware classes and families. For packed malware programs, only 2.9\% of anti-virus products in VirusTotal correctly recognize their classes and families, while for unpacked programs, this figure rises to 17\%. Alarmingly, half of the packed malware categories are classified with an accuracy below 1\%, compared to only 13.8\% for unpacked categories. This stark difference suggests that many anti-virus products in VirusTotal do not unpack programs before classification, significantly reducing their ability to accurately classify malware programs. This highlights the critical importance of unpacking in achieving reliable malware classification and underscores a major limitation in existing commercial antivirus solutions.

The approach proposed by Adkins \emph{et al.} \cite{adkins2013heuristic} stands out with an accuracy of 99\% since instruction replacement still preserves the similarity among malware instances. However, it introduces more false positives due to instruction abstraction. Additionally, the method used by Chen \emph{et al.} \cite{chen2020malware} performs worse than \cite{adkins2013heuristic}. It loses more critical information needed to distinguish different malware classes and families. Furthermore, for malware programs in the same category, even a modest change to the instructions and instruction order significantly changes hash values, leading to a poorer performance. In contrast, our framework, \textit{VisUnpack}, achieves the highest accuracy, precision, recall, and F1 score. This demonstrates the effectiveness of considering local similarities within basic blocks as well as their correlations across different scales in detecting malware. This approach proves to be a reliable method for malware classification, showing promising potential in thwarting attacks. Compared to the methods that individually apply VGG11 to gray images (VisUnpack\_grey) and feature enhancement (VisUnpack\_fe), our approach has similar performance. However, VisUnpack can save more than 97\% space and 99.1\% than VisUnpack\_grey and VisUnpack\_fe respectively. \textit{VisUpack} has a higher accuracy than local similarity descriptors (VisUnpack\_ssd).  This indicates that the combination of local similarity calculation and mutual relationship enhancement can preserve and enhance similarity while achieving significant space savings.


   	
   	

\subsection{Time and Resource Consumption}
Given the expenses related to storage and computing resources, it's economically unfeasible to keep original samples for retraining models to adapt to future changes. Additionally, considering the time-sensitive aspect of malware classification within antivirus systems, it's crucial that the process of identifying malware programs doesn't incur significant delays, which is typically associated with dynamic analysis. To evaluate the efficiency of our proposed framework, we measured the CPU time required for our dataset to pass through \textit{VisUnpack} during the test phase, as shown in Equation \eqref{eq:efficiency}. Our framework is designed for practical deployment without the need for server infrastructure or GPU utilization. To assess resource consumption, we consider the size of the dataset during both training and testing phases of machine learning models.

In our comparative analysis of malware classification methods using the Malfiner dataset, we evaluated several recent studies, including \cite{Nataraj}, \cite{adkins2013heuristic}, \cite{Saxe}, \cite{Yakura}, \cite{cui2018detection}, \cite{Xusheng}, \cite{liu2019new}, \cite{Xiang}, \cite{Ouahab}, \cite{Baoguo}, \cite{chen2020malware}, and \cite{Malbrain}. As shown in Table \ref{tab:efficiency1}, all methods take longer to process the packed dataset compared to the unpacked one due to the larger size of packed dataset. \textit{VisUnpack}, requiring 11.2 seconds per sample on packed data and 9.7 seconds on unpacked data, is at least 94.8\% faster than methods like \cite{chen2020malware} and \cite{adkins2013heuristic} that also use basic blocks as comparison units, while also saving over 71\% in storage. Although VisUnpack takes longer to classify samples compared to methods such as DNN \cite{Saxe}, KNN \cite{Ouahab}, or CNN \cite{Xusheng}, it achieves significant storage savings, exceeding 94\%, and offers much higher accuracy. While VisUnpack\_ssd achieves shorter processing times and reduced storage, it falls short in accuracy compared to VisUnpack, highlighting a trade-off between time, storage, and precision. These results demonstrate that VisUnpack is a highly precise and storage-efficient solution, making it ideal for large-scale malware classification tasks.

	\begin{table}
		\centering
		\caption{Time and Resource Consumption on Packed/Unpacked Malfiner}
		\begin{tabular}{|c|c|c|}
			\hline
			& avg\_time & Storage\\
			\hline
			  DNN \cite{Saxe} & 26.2ms/22.7ms& 18.1GB/13.5GB\\
			\hline
            KNN  \cite{Ouahab}& 21.1ms/20.3ms&18.1GB/13.5GB\\            			\hline
            CNN \cite{Xusheng}&25.7ms/23.8ms &18.1GB/13.5GB\\
            \hline
            Autoencoder \cite{Xiang}&25.4ms/23.8ms&18.1GB/13.5GB\\
            \hline
            CNN\_Attention \cite{Yakura}&25.1ms/22.8ms&18.1GB/13.5GB\\
            \hline
            DRBA \cite{Zhihua}&26.8ms/23.9ms&18.1GB/13.5GB\\
            \hline
            GIST  \cite{Nataraj} &21.08ms/19.38ms  &18.1GB/13.5GB\\
            \hline
            VisMal  \cite{Malbrain} & 28ms/25.1ms &20GB/16GB\\
            \hline
		MDMC \cite{Baoguo}& 5.7ms/8.2ms &6.6GB/6.6GB\\
			\hline
            LBP \cite{liu2019new}&344.18ms/284.1ms&18.1GB/13.5GB\\
            \hline
            Chen \emph{et al.} \cite{chen2020malware} & 26.7s/18.9s &0.8GB/1.38GB\\
            \hline
            Adkins \emph{et al.} \cite{adkins2013heuristic} & 29.4s/21.1s &222.1GB/222.1GB\\
            \hline
            VisUnpack\_grey & 27.4ms/24.3ms & 18.1GB/13.5GB \\
            \hline
            VisUnpack\_ssd & 11.1s/9.7s &0.3GB/0.3GB \\
            \hline
            VisUnpack\_fe &62.3ms/51ms &59.2GB/46.9GB \\
            \hline
            VisUnpack& 11.2s/9.7s &0.5GB/0.4GB \\
            \hline
		\end{tabular}
		\label{tab:efficiency1}
	\end{table}

\section{Conclusion}
\label{sec:conclusion}
In this paper, we introduce a novel malware dataset, Malfiner, comprising 27,106 samples collected over eight years. This dataset provides a granular level of categorization, distinguishing both malware classes and families, and has been rigorously validated using over 70 third-party anti-virus products on VirusTotal, as well as through dynamic analysis and reverse engineering. As such, Malfiner offers a comprehensive reflection of contemporary malware attack trends and facilitates the development of automated recovery techniques. Additionally, we present \textit{VisUnpack}, a context-sensitive malware visualization classification framework that distinguishes between malware programs, even those within the same class but different families, outperforming most anti-virus products on VirusTotal and existing works. Our experiments on packed and unpacked Malfiner datasets reveal the detrimental impact of packing on malware classification accuracy and highlight that many VirusTotal anti-virus products fail to handle packed samples effectively. These findings underscore the necessity of unpacking malware samples before classification. Furthermore, \textit{VisUnpack} significantly reduces the computational overhead of model training while preserving critical similarities among malware programs within the same class and family. Moreover, our approach differs from many existing methods, which often focus on statistical, correlation-based, opcode or operand abstraction techniques for malware classification. Instead, our method returns to the original concept of malware classification, focusing on detecting similarities between malware programs. It captures local semantics and resemblances within basic blocks, as well as the global correlations between them. By focusing on basic blocks as the fundamental unit of program functionality, our method achieves greater resilience to changes and exhibits enhanced robustness compared to approaches relying on global statistical distributions or bytecode correlations.

	
	%

	\ifCLASSOPTIONcaptionsoff
	\newpage
	\fi

	
	
	%
\bibliographystyle{IEEEtran}
\bibliography{sample}

\begin{thebibliography}{10}
\providecommand{\url}[1]{#1}
\csname url@samestyle\endcsname
\providecommand{\newblock}{\relax}
\providecommand{\bibinfo}[2]{#2}
\providecommand{\BIBentrySTDinterwordspacing}{\spaceskip=0pt\relax}
\providecommand{\BIBentryALTinterwordstretchfactor}{4}
\providecommand{\BIBentryALTinterwordspacing}{\spaceskip=\fontdimen2\font plus
\BIBentryALTinterwordstretchfactor\fontdimen3\font minus \fontdimen4\font\relax}
\providecommand{\BIBforeignlanguage}[2]{{%
\expandafter\ifx\csname l@#1\endcsname\relax
\typeout{** WARNING: IEEEtran.bst: No hyphenation pattern has been}%
\typeout{** loaded for the language `#1'. Using the pattern for}%
\typeout{** the default language instead.}%
\else
\language=\csname l@#1\endcsname
\fi
#2}}
\providecommand{\BIBdecl}{\relax}
\BIBdecl

\bibitem{anti-test}
D.~Delić, ``Malware statistics and facts in 2024 – how to best protect your business and yourself?'' \emph{CYBERCRIME MAGAZINE}, 15 March 2022, available at: \url{https://proprivacy.com/blog/malware-statistics-and-facts-2023-how-to-protect-yourself}.

\bibitem{mihai}
M.~Christodorescu and S.~Jha, ``Static analysis of executables to detect malicious patterns,'' in \emph{12th USENIX Security Symposium (USENIX Security 03)}.\hskip 1em plus 0.5em minus 0.4em\relax Washington, D.C.: USENIX Association, aug 2003.

\bibitem{Simone}
S.~Aonzo, Y.~Han, A.~Mantovani, and D.~Balzarotti, ``Humans vs. machines in malware classification,'' in \emph{32nd USENIX Security Symposium (USENIX Security 23)}.\hskip 1em plus 0.5em minus 0.4em\relax Anaheim, CA: USENIX Association, aug 2023, pp. 1145--1162.

\bibitem{Korczynski}
\BIBentryALTinterwordspacing
D.~Korczynski and H.~Yin, ``Capturing malware propagations with code injections and code-reuse attacks,'' in \emph{Proceedings of the 2017 ACM SIGSAC Conference on Computer and Communications Security}, ser. CCS '17.\hskip 1em plus 0.5em minus 0.4em\relax New York, NY, USA: Association for Computing Machinery, 2017, p. 1691–1708. [Online]. Available: \url{https://doi.org/10.1145/3133956.3134099}
\BIBentrySTDinterwordspacing

\bibitem{gu2007bothunter}
G.~Gu, P.~A. Porras, V.~Yegneswaran, M.~W. Fong, and W.~Lee, ``Bothunter: Detecting malware infection through ids-driven dialog correlation.'' in \emph{USENIX Security Symposium}, vol.~7, 2007, pp. 1--16.

\bibitem{peng2014x}
F.~Peng, Z.~Deng, X.~Zhang, D.~Xu, Z.~Lin, and Z.~Su, ``$\{$X-Force$\}$:$\{$Force-Executing$\}$ binary programs for security applications,'' in \emph{23rd USENIX Security Symposium (USENIX Security 14)}, 2014, pp. 829--844.

\bibitem{eom2024r2i}
H.~Eom, D.~Kim, S.~Lim, H.~Koo, and S.~Hwang, ``R2i: A relative readability metric for decompiled code,'' \emph{Proceedings of the ACM on Software Engineering}, vol.~1, no. FSE, pp. 383--405, 2024.

\bibitem{shoshitaishvili2016state}
Y.~Shoshitaishvili, R.~Wang, C.~Salls, N.~Stephens, M.~Polino, A.~Dutcher, J.~Grosen, S.~Feng, C.~Hauser, C.~Kruegel, and G.~Vigna, ``{SoK: (State of) The Art of War: Offensive Techniques in Binary Analysis},'' in \emph{IEEE Symposium on Security and Privacy}, 2016.

\bibitem{SAVIOR}
Y.~Chen, P.~Li, J.~Xu, S.~Guo, R.~Zhou, Y.~Zhang, T.~Wei, and L.~Lu, ``Savior: Towards bug-driven hybrid testing,'' in \emph{2020 IEEE Symposium on Security and Privacy (SP)}, 2020, pp. 1580--1596.

\bibitem{ConcolicTesting}
J.~Choi, J.~Jang, C.~Han, and S.~K. Cha, ``Grey-box concolic testing on binary code,'' in \emph{2019 IEEE/ACM 41st International Conference on Software Engineering (ICSE)}, 2019, pp. 736--747.

\bibitem{Jiancong}
\BIBentryALTinterwordspacing
L.~Cui, J.~Cui, Y.~Ji, Z.~Hao, L.~Li, and Z.~Ding, ``Api2vec: Learning representations of api sequences for malware detection,'' in \emph{Proceedings of the 32nd ACM SIGSOFT International Symposium on Software Testing and Analysis}, ser. ISSTA 2023.\hskip 1em plus 0.5em minus 0.4em\relax New York, NY, USA: Association for Computing Machinery, 2023, p. 261–273. [Online]. Available: \url{https://doi.org/10.1145/3597926.3598054}
\BIBentrySTDinterwordspacing

\bibitem{Cuiying}
\BIBentryALTinterwordspacing
C.~Gao, G.~Huang, H.~Li, B.~Wu, Y.~Wu, and W.~Yuan, ``A comprehensive study of learning-based android malware detectors under challenging environments,'' in \emph{Proceedings of the 46th IEEE/ACM International Conference on Software Engineering}, ser. ICSE '24.\hskip 1em plus 0.5em minus 0.4em\relax New York, NY, USA: Association for Computing Machinery, 2024. [Online]. Available: \url{https://doi.org/10.1145/3597503.3623320}
\BIBentrySTDinterwordspacing

\bibitem{Xiang}
X.~Jin, X.~Xing, H.~Elahi, G.~Wang, and H.~Jiang, ``A malware detection approach using malware images and autoencoders,'' in \emph{2020 IEEE 17th International Conference on Mobile Ad Hoc and Sensor Systems (MASS)}, 2020, pp. 1--6.

\bibitem{Xusheng}
X.~Xiao and S.~Yang, ``An image-inspired and cnn-based android malware detection approach,'' in \emph{2019 34th IEEE/ACM International Conference on Automated Software Engineering (ASE)}, 2019, pp. 1259--1261.

\bibitem{virustotal}
\BIBentryALTinterwordspacing
G.~Darroch, ``How it works,'' \emph{VirusTotal}, 2024. [Online]. Available: \url{https://docs.virustotal.com/docs/how-it-works}
\BIBentrySTDinterwordspacing

\bibitem{vieira2010cohen}
S.~M. Vieira, U.~Kaymak, and J.~M. Sousa, ``Cohen's kappa coefficient as a performance measure for feature selection,'' in \emph{International conference on fuzzy systems}.\hskip 1em plus 0.5em minus 0.4em\relax IEEE, 2010, pp. 1--8.

\bibitem{dataset}
\BIBentryALTinterwordspacing
X.~Ugarte-Pedrero, M.~Graziano, and D.~Balzarotti, ``A close look at a daily dataset of malware samples,'' \emph{ACM Trans. Priv. Secur.}, vol.~22, no.~1, jan 2019. [Online]. Available: \url{https://doi.org/10.1145/3291061}
\BIBentrySTDinterwordspacing

\bibitem{Zhihua}
Z.~Cui, F.~Xue, X.~Cai, Y.~Cao, G.-g. Wang, and J.~Chen, ``Detection of malicious code variants based on deep learning,'' \emph{IEEE Transactions on Industrial Informatics}, vol.~14, no.~7, pp. 3187--3196, 2018.

\bibitem{Malbrain}
F.~Zhong, Z.~Chen, M.~Xu, G.~Zhang, D.~Yu, and X.~Cheng, ``Malware-on-the-brain: Illuminating malware byte codes with images for malware classification,'' \emph{IEEE Transactions on Computers}, vol.~72, no.~2, pp. 438--451, 2023.

\bibitem{Nataraj}
\BIBentryALTinterwordspacing
L.~Nataraj, S.~Karthikeyan, G.~Jacob, and B.~S. Manjunath, ``Malware images: visualization and automatic classification,'' in \emph{Proceedings of the 8th International Symposium on Visualization for Cyber Security}, ser. VizSec '11.\hskip 1em plus 0.5em minus 0.4em\relax New York, NY, USA: Association for Computing Machinery, 2011. [Online]. Available: \url{https://doi.org/10.1145/2016904.2016908}
\BIBentrySTDinterwordspacing

\bibitem{Torralba}
Torralba, Murphy, Freeman, and Rubin, ``Context-based vision system for place and object recognition,'' in \emph{Proceedings Ninth IEEE International Conference on Computer Vision}, 2003, pp. 273--280 vol.1.

\bibitem{Aude}
A.~T. Aude~Oliva, ``Modeling the shape of the scene: A holistic representation of the spatial envelope,'' \emph{International Journal of Computer Vision}, vol.~42, no.~3, pp. 145--175, 2001.

\bibitem{Makandar}
\BIBentryALTinterwordspacing
A.~Makandar and A.~Patrot, ``Wavelet statistical feature based malware class recognition and classification using supervised learning classifier,'' \emph{Oriental journal of computer science and technology}, vol.~10, pp. 400--406, 2017. [Online]. Available: \url{https://www.computerscijournal.org/pdf/vol10no2/OJCST_Vol10_N2_p_400-407.pdf}
\BIBentrySTDinterwordspacing

\bibitem{Baoguo}
\BIBentryALTinterwordspacing
B.~Yuan, J.~Wang, D.~Liu, W.~Guo, P.~Wu, and X.~Bao, ``Byte-level malware classification based on markov images and deep learning,'' \emph{Computers \& Security}, vol.~92, p. 101740, 2020. [Online]. Available: \url{https://www.sciencedirect.com/science/article/pii/S0167404820300262}
\BIBentrySTDinterwordspacing

\bibitem{liu2019new}
Y.-s. Liu, Y.-K. Lai, Z.-H. Wang, and H.-B. Yan, ``A new learning approach to malware classification using discriminative feature extraction,'' \emph{IEEE Access}, vol.~7, pp. 13\,015--13\,023, 2019.

\bibitem{chen2020malware}
J.~Chen, ``A malware classification method based on basic block and cnn,'' in \emph{Neural Information Processing: 27th International Conference, ICONIP 2020, Bangkok, Thailand, November 18--22, 2020, Proceedings, Part IV 27}.\hskip 1em plus 0.5em minus 0.4em\relax Springer, 2020, pp. 275--283.

\bibitem{adkins2013heuristic}
F.~Adkins, L.~Jones, M.~Carlisle, and J.~Upchurch, ``Heuristic malware detection via basic block comparison,'' in \emph{2013 8th International Conference on Malicious and Unwanted Software:" The Americas"(MALWARE)}.\hskip 1em plus 0.5em minus 0.4em\relax IEEE, 2013, pp. 11--18.

\bibitem{Saxe}
J.~Saxe and K.~Berlin, ``Deep neural network based malware detection using two dimensional binary program features,'' in \emph{2015 10th International Conference on Malicious and Unwanted Software (MALWARE)}, 2015, pp. 11--20.

\bibitem{Ouahab}
I.~Ben Abdel~Ouahab, M.~Bouhorma, A.~A. Boudhir, and L.~El~Aachak, ``Classification of grayscale malware images using the k-nearest neighbor algorithm,'' in \emph{Innovations in Smart Cities Applications Edition 3}, M.~Ben~Ahmed, A.~A. Boudhir, D.~Santos, M.~El~Aroussi, and {\.{I}}.~R. Karas, Eds.\hskip 1em plus 0.5em minus 0.4em\relax Cham: Springer International Publishing, 2020, pp. 1038--1050.

\bibitem{Yakura}
\BIBentryALTinterwordspacing
H.~Yakura, S.~Shinozaki, R.~Nishimura, Y.~Oyama, and J.~Sakuma, ``Malware analysis of imaged binary samples by convolutional neural network with attention mechanism,'' ser. CODASPY '18.\hskip 1em plus 0.5em minus 0.4em\relax New York, NY, USA: Association for Computing Machinery, 2018, p. 127–134. [Online]. Available: \url{https://doi.org/10.1145/3176258.3176335}
\BIBentrySTDinterwordspacing

\bibitem{peid}
``Detect packers on pe files using signatures.'' \url{https://github.com/packing-box/peid}, accessed: 2024-12-30.

\bibitem{zhong2024enhancing}
F.~Zhong, Q.~Hu, Y.~Jiang, J.~Huang, C.~Zhang, and D.~Wu, ``Enhancing malware classification via self-similarity techniques,'' \emph{IEEE Transactions on Information Forensics and Security}, 2024.

\bibitem{Malpedia}
``Malpedia is a free service offered by fraunhofer fkie.'' \url{https://malpedia.caad.fkie.fraunhofer.de/}, accessed: 2024-12-28.

\bibitem{simonyan2014very}
K.~Simonyan and A.~Zisserman, ``Very deep convolutional networks for large-scale image recognition,'' \emph{International Conference on Learning Representations 2014}, 2014.

\bibitem{upx}
``the ultimate packer for executables,'' \url{https://upx.github.io/}, accessed: 2024-12-30.

\bibitem{petite}
``Win32 executable compressor,'' \url{https://www.un4seen.com/petite/ }, accessed: 2024-12-28.

\bibitem{PECompact}
``Pecompact – windows (pe) executable compressor,'' \url{https://bitsum.com/portfolio/pecompact/}, accessed: 2024-12-30.

\bibitem{mpress}
``Mpress is a free, high-performance executable packer for pe32/pe32+/.net/mac-darwin executable formats!'' \url{https://www.autohotkey.com/mpress/mpress_web.htm }, accessed: 2024-12-30.

\bibitem{unipacker}
``Unpacking pe files using unicorn engine,'' \url{https://github.com/unipacker/unipacker }, accessed: 2024-12-28.

\bibitem{unpacme}
``Automated malware unpacking and artifact extraction,'' \url{https://www.unpac.me/ }, accessed: 2024-12-28.

\bibitem{cuckoo}
D.~Oktavianto and I.~Muhardianto, \emph{Cuckoo malware analysis}.\hskip 1em plus 0.5em minus 0.4em\relax Packt Publishing Ltd, 2013.

\bibitem{Calleja}
A.~Calleja, J.~Tapiador, and J.~Caballero, ``The malsource dataset: Quantifying complexity and code reuse in malware development,'' \emph{IEEE Transactions on Information Forensics and Security}, vol.~14, no.~12, pp. 3175--3190, 2019.

\bibitem{Xiaozhu}
\BIBentryALTinterwordspacing
X.~Meng and B.~P. Miller, ``Binary code is not easy,'' in \emph{Proceedings of the 25th International Symposium on Software Testing and Analysis}, ser. ISSTA 2016.\hskip 1em plus 0.5em minus 0.4em\relax New York, NY, USA: Association for Computing Machinery, 2016, p. 24–35. [Online]. Available: \url{https://doi.org/10.1145/2931037.2931047}
\BIBentrySTDinterwordspacing

\bibitem{Andriesse}
\BIBentryALTinterwordspacing
D.~Andriesse, X.~Chen, V.~van~der Veen, A.~Slowinska, and H.~Bos, ``An {In-Depth} analysis of disassembly on {Full-Scale} x86/x64 binaries,'' in \emph{25th USENIX Security Symposium (USENIX Security 16)}.\hskip 1em plus 0.5em minus 0.4em\relax Austin, TX: USENIX Association, Aug. 2016, pp. 583--600. [Online]. Available: \url{https://www.usenix.org/conference/usenixsecurity16/technical-sessions/presentation/andriesse}
\BIBentrySTDinterwordspacing

\bibitem{clahe}
K.~Zuiderveld, ``Contrast limited adaptive histogram equalization,'' \url{http://cas.xav.free.fr/Graphics\%20Gems\%204\%20-\%20Paul\%20S.\%20Heckbert.pdf}, 1993.

\bibitem{belongie2002shape}
S.~Belongie, J.~Malik, and J.~Puzicha, ``Shape matching and object recognition using shape contexts,'' \emph{IEEE transactions on pattern analysis and machine intelligence}, vol.~24, no.~4, pp. 509--522, 2002.

\bibitem{equalization}
R.~Dorothy, R.~Joany, R.~J. Rathish, S.~S. Prabha, and S.~rajendran, ``Image enhancement by histogram equalization,'' \emph{Int. J. Nano. Corr. Sci. Engg}, vol.~2, no.~4, pp. 21--30, 2015.

\bibitem{enhancement}
X.~Yu, C.~Kang, D.~S. Guttery, S.~Kadry, Y.~Chen, and Y.-D. Zhang, ``Resnet-scda-50 for breast abnormality classification,'' \emph{IEEE/ACM Transactions on Computational Biology and Bioinformatics}, vol.~18, no.~1, pp. 94--102, 2021.

\bibitem{ahe}
``Adaptive histogram equalization,'' \url{https://en.wikipedia.org/wiki/ Adaptivehistogramequalization citenote-clahe87-3}, 2021.

\bibitem{virusshare}
``Virusshare.com - because sharing is caring,'' \url{https://virusshare.com/}, 2023, accessed: 2023-2-14.

\bibitem{classguide}
``Malware classification guide,'' \url{https://any.run/cybersecurity-blog/malware-classification-guide/ }, 2020.

\bibitem{cui2018detection}
Z.~Cui, F.~Xue, X.~Cai, Y.~Cao, G.-g. Wang, and J.~Chen, ``Detection of malicious code variants based on deep learning,'' \emph{IEEE Transactions on Industrial Informatics}, vol.~14, no.~7, pp. 3187--3196, 2018.

\end{thebibliography}

\begin{IEEEbiography}[{\includegraphics[width=1in,height=1.10in,clip]{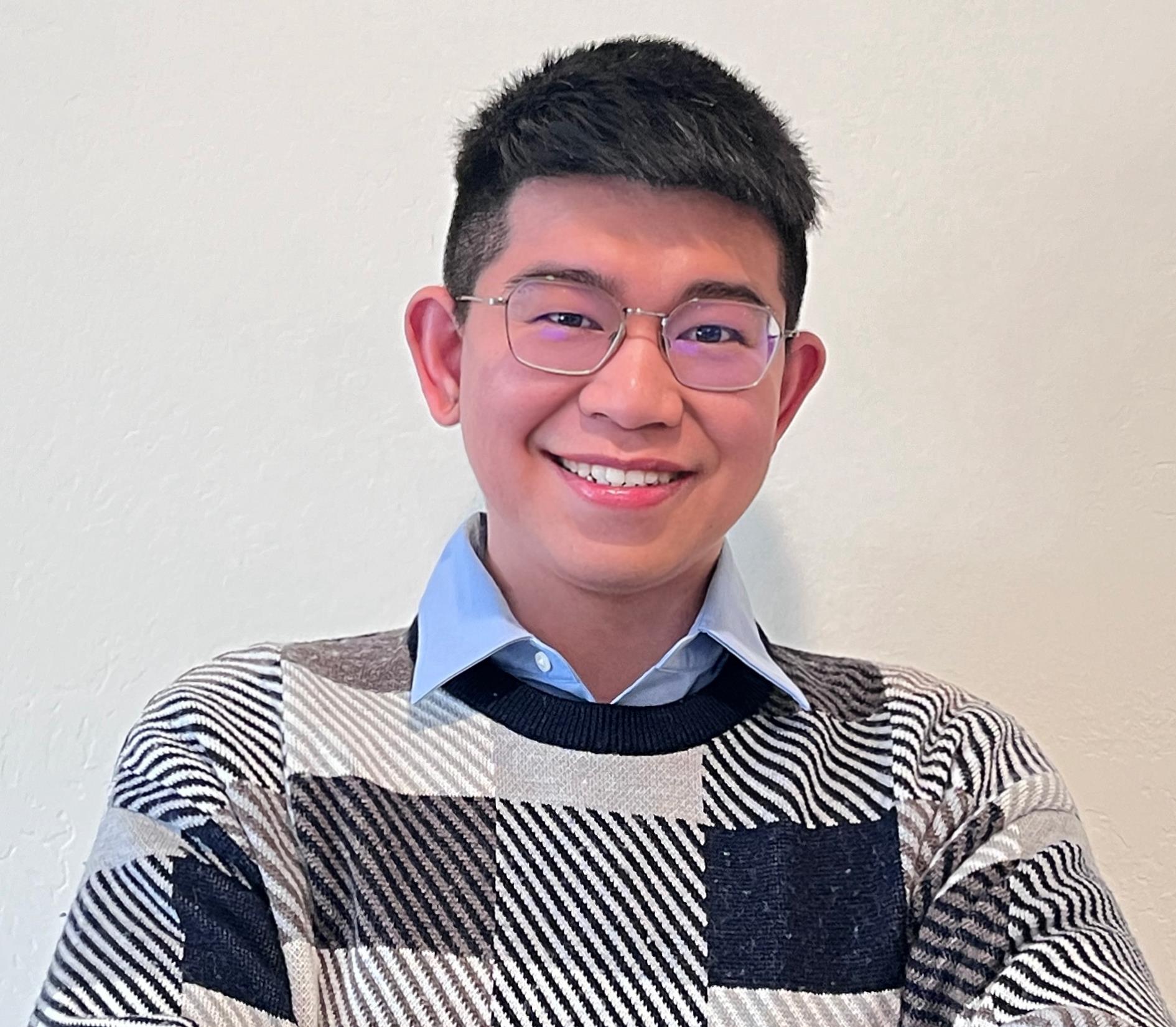}}]{Fangtian Zhong} received his Ph.D. degree in Computer Science from the George Washington University in 2021. After that, he was a Post-Doctoral Scholar with The Pennsylvania State University and the University of Notre Dame, respectively. He is an Assistant Professor with Montana State University. His research focuses on software security, program analysis, and machine learning for cybersecurity.
\end{IEEEbiography}

\begin{IEEEbiography}[{\includegraphics[width=1in,height=1.32in,clip,keepaspectratio]{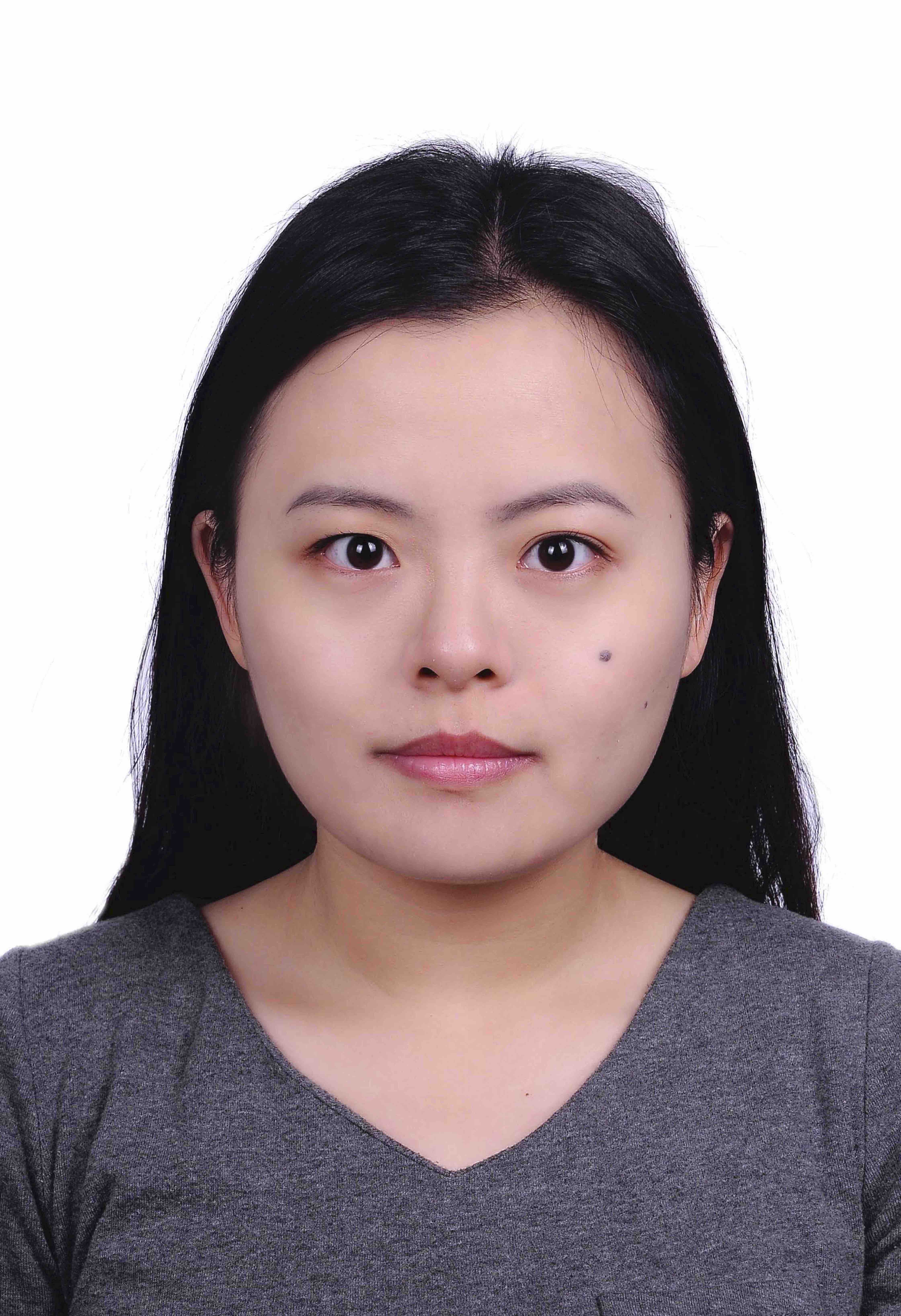}}] {Qin Hu} received her Ph.D. degree in Computer Science from the George Washington University in 2019. She is currently an assistant professor with the Department of Computer Science at Georgia State University. She has served as the Editor/Guest Editor for several journals, the TPC/Publicity Co-chair for several workshops, and the TPC Member for several international conferences. Her research interests include wireless and mobile security, edge computing, blockchain, and federated learning.

\end{IEEEbiography}

\begin{IEEEbiography}[{\includegraphics[width=1in,height=1.25in,clip,keepaspectratio]{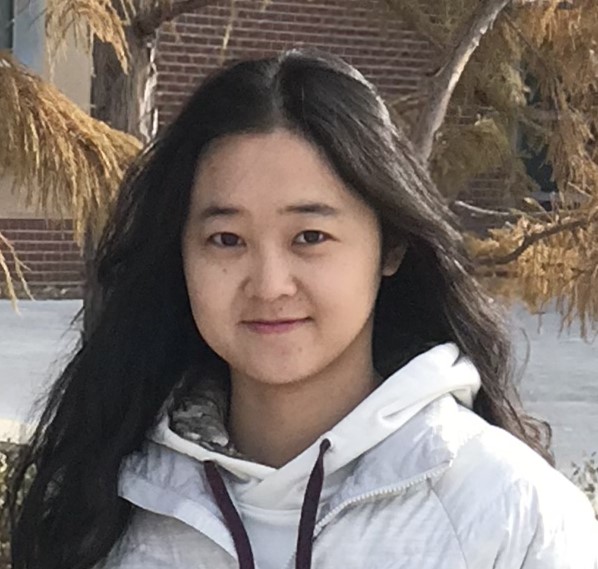}}]{Yili Jiang}  received her Ph.D. degree in Computer Engineering from the University of Nebraska-Lincoln, NE, USA, in 2022. She is currently an assistant professor in the Department of Computer Science at Georgia State University, GA, USA. Her research interests include cybersecurity, machine learning, cloud/edge computing, and wireless networks.
\end{IEEEbiography}

\begin{IEEEbiography}[{\includegraphics[width=1in,height=1.25in,clip,keepaspectratio]{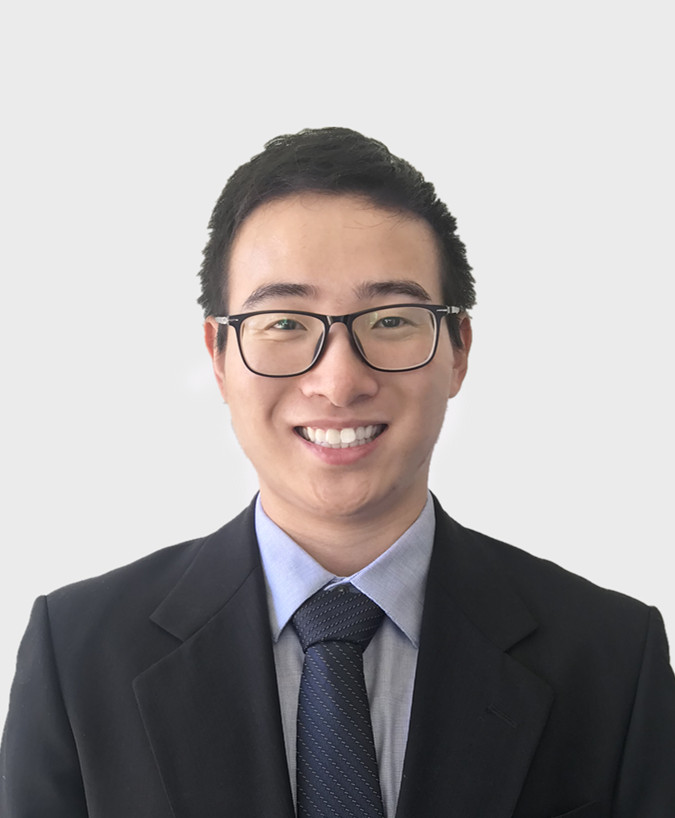}}]{Jiaqi Huang} received his Ph.D. in Computer Engineering from the University of Nebraska-Lincoln, NE, USA, in 2020. He is currently an assistant professor in the Department of Computer Science and Cybersecurity at the University of Central Missouri, MO, USA. His research interests include cybersecurity, applied cryptography, machine learning, connected autonomous vehicles, and wireless networks.
\end{IEEEbiography}

\begin{IEEEbiography}[{\includegraphics[width=1in,height=1.25in,clip,keepaspectratio]{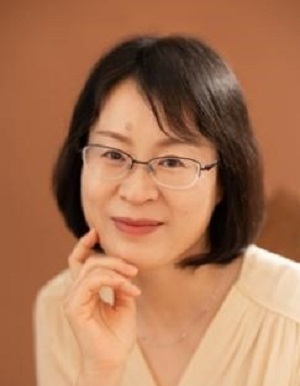}}]{Xiuzhen Cheng}received her MS and Ph.D. degrees in computer science from the University of Minnesota – Twin Cities, in 2000 and 2002, respectively. She was a faculty member at the Department of Computer Science, The George Washington University, from 2002 to 2020. Currently, she is a professor of computer science at Shandong University, Qingdao, China. Her research focuses on blockchain computing, IoT Security, and privacy-aware computing. She is a Fellow of IEEE, CSEE and AAIA.

\end{IEEEbiography}	
\end{document}